\documentclass[useAMS,usenatbib]{mn2e}
\usepackage{epsfig}
\usepackage{amsmath}
\usepackage{amssymb}
\usepackage{footmisc}
\newcommand{\hst}{\textit{HST}}

\newcommand{\acsb}{\hbox{$B_{435}$}}
\newcommand{\acsv}{\hbox{$V_{606}$}}
\newcommand{\acsi}{\hbox{$i_{775}$}}
\newcommand{\acsz}{\hbox{$z_{850}$}}

\newcommand{\ha}{\hbox{H$\alpha$}}

\newcommand{\lsim}{\lesssim}
\newcommand{\gsim}{\gtrsim}

\newcommand{\eg}{e.g.}

\newcommand{\msol}{\hbox{$M_\odot$}}

\newcommand{\zsol}{\hbox{$Z_\odot$}}

\newcommand{\infinity}{\hbox{$\infty$}}

\newcommand{\muv}{\hbox{$M_\mathrm{UV}^\prime$}}
\newcommand{\mstar}{\hbox{$M_\ast^\prime$}}
\newcommand{\mgas}{\hbox{$M_\mathrm{gas}^\prime$}}

\title[Rising Star-Formation Histories and Gas Accretion of Distant Galaxies]{The Rising Star-Formation Histories of Distant Galaxies and Implications for Gas Accretion with Time}
\author[C. Papovich et al.]{Casey Papovich$^{1,2}$\thanks{E-mail:
papovich@physics.tamu.edu}, Steven L. Finkelstein$^{1,2}$, Henry C. Ferguson$^{3}$,
\newauthor
Jennifer M. Lotz$^{3,4}$\thanks{Leo Goldberg Fellow}, 
and Mauro Giavalisco$^{5}$ \\
$^{1}$Department of Physics and Astronomy, Texas A\&M University, College Station, TX 77845-4242, USA\\
$^{2}$George P. and Cynthia Woods Mitchell Institute for Fundamental Physics and Astronomy \\
$^{3}$Space Telescope Science Institute, 3700 San Martin Drive, Baltimore, MD 21218, USA \\
$^{4}$National Optical Astronomy Observatory, 950 N. Cherry Avenue,
Tucson, AZ 85719, USA \\
$^{5}$Department of Astronomy, University of Massachusetts, Amherst,
MA 01003, USA
}
\begin{document}

\date{Accepted 2010 November 1.  Received 2010 October 18; in original form 2010 July 12}

\pagerange{\pageref{firstpage}--\pageref{lastpage}} \pubyear{2010}

\maketitle

\label{firstpage}

\begin{abstract}  Distant galaxies show correlations
between their current star-formation rates (SFRs) and stellar masses,
implying that their star-formation histories (SFHs) are highly
similar.  Moreover, observations show that the UV luminosities and
stellar masses grow from z=8 to z=3, implying that the SFRs increase
with time.  We compare the cosmologically averaged evolution in
galaxies at $3 < z < 8$ at constant comoving number density, $n = 2
\times 10^{-4}$ Mpc$^{-3}$.  This allows us to study the evolution of
stellar mass and star formation in the galaxy predecessors and
descendants in ways not possible using galaxies selected at constant
stellar mass or SFR, quantities that evolve strongly in time.  We show
that the cosmologically averaged SFRs of these galaxies increase
smoothly from $z=8$ to 3 as $\Psi(t) \sim t^{\alpha}$ with $\alpha =
1.7 \pm 0.2$.  This conflicts with assumptions that the SFR
is either constant or declines exponentially in time.  Furthermore, we
show that the stellar mass growth in these galaxies is consistent with
this derived SFH.  This provides evidence that the slope of the
high-mass end of the IMF is approximately Salpeter unless the duty
cycle of star formation is much less than unity.   We argue that these
relations follow from gas accretion (either through accretion or
delivered by mergers) coupled with galaxy disk growth under the
assumption that the SFR depends on the local gas surface density. This
predicts that  gas fractions decrease from $z=8$ to 3 on average
as $f_\mathrm{gas} \sim (1 + z)^{0.9}$ for galaxies with this number
density.  The implied galaxy gas accretion rates at $z > 4$ are
as fast and may even exceed the SFR: this is the ``gas accretion
epoch''.  At $z < 4$ the SFR overtakes the implied gas accretion rate,
indicating a period where galaxies consume gas faster than it is
acquired.  At $z \lsim 3$, galaxies with this number density depart
from these relations implying that star formation and gas accretion
are slowed at later times.
\end{abstract}


\section{INTRODUCTION}

The process by which galaxies acquire their baryonic gas and convert
it to stars is one of the main questions for theories of galaxy
formation.  However, it has been very difficult to constrain directly
the evolution of these  processes primarily because it is challenging
to link the descendants and progenitors of galaxies at
different snapshots in redshift.     Currently, we have only a weak
observational understanding of the  star-formation histories (SFHs) of
distant galaxies, which is arguably one of the most important
quantities to measure because it governs directly how rapidly galaxies
build up their stellar content.  

Nevertheless, there are recent hints from the interpretation of
observational data that distant galaxies experience SFHs where the SFR
increases with time (decreasing redshift).    The relatively tight
correlation and low scatter between the star formation rates (SFRs)
and stellar masses in distant galaxies
\citep[\eg,][]{daddi07a,noeske07a,stark09,magdis10} imply that the
majority of galaxies at high redshift sustain high levels of star
formation for prolonged time-scales of $\sim$0.5--1~Gyr, with high
``duty cycles''.\footnote{Here we take the \textit{duty cycle} to be
the ratio of the time duration of star-formation in galaxies relative
to the total time interval probed by the galaxy survey, see, e.g.,
\citet{kslee09}.}  Therefore short-lived starbursts driven by major
mergers likely play a minor role.  If the SFRs of these galaxies have been
constant or declining in time, then it implies that the progenitors of
these galaxies would have had higher specific SFRs (SFRs per unit
stellar mass).  However, \citet{stark09} observed a lack of evolution
in the specific SFRs of UV-selected galaxies from $z=4-6$, which
excludes the possibility that galaxies at $z=4$ have been forming
stars at constant or declining rates since $z\gsim 5$.
%
%
\citet{fink10} noted
that both the characteristic luminosity of the UV luminosity function,
$L_\mathrm{UV}^\ast$, and the stellar mass of galaxies at the
characteristic luminosity increase with decreasing redshift from
$z\sim 8$ to 3, which is most easily explainable if distant galaxies
experienced SFRs that increase with time.   Furthermore, recent work
by \citet{mara10} argue that rising SFHs better reproduce the observed
colours of galaxies at $z\sim 2$ compared to models that decline with
time \citep[see also,][]{sklee10}. 

Support for periods of extended star formation, such as that
expected from a rising or constant SFH,  also comes from kinematic
measurements of many star-forming galaxies at $z\sim 1.5$--4, which
show evidence for extended rotating gaseous disks not dominated by
spheroids or mergers
\citep[\eg,][]{genzel08,daddi09a,daddi09b,forster09,law09,wright09,cari10,tacc10}. The
existence of gaseous disks likely requires that distant galaxies form
stars at a quasi-steady pace, and not in short periods of
enhanced star formation such as merger-driven starbursts
\citep[see,][]{elme08,bour08,dekel09b,renz09}.  Comparing galaxy
SFRs to their gas masses yields gas consumption
time-scales of a few hundred Myr, typically much shorter than a Hubble
time at the observed redshift.  In order for the galaxies to maintain
their high levels of star formation, it follows that they must receive
cold gas from the intergalactic medium (IGM) to replenish their
supplies of fresh gas to fuel the continued star formation
\citep[\eg,][]{erb08,proc09}.   

SFRs that increase with time for $z \gsim 3$ are an outcome from
hydrodynamic and semi-analytic simulations.  \citet{mara10}  favor
SFRs that increase with time in part because these models reproduce
the SFRs and stellar masses of galaxies from their semi-analytic
model.   Similar conclusions are obtained by \citet{sklee10} using
SFHs from a semi-analytic model based on \citet{some01}.    Using a
cosmological hydrodynamic simulation, \citet{finl07} predict that
galaxies at $z\ge 6$ experience SFRs that increase with time, and they
show that these SFHs reproduce the UV luminosity functions and
observed colours of distant galaxies \citep{finl06,finl07,finl10}.
Interestingly, the predicted SFHs in these simulations are
scale-invariant, varying in mass only by a scale factor, implying that
measurements of the SFHs of distant galaxies have strong constraining
power on models.  In the simulations, these smoothly-rising SFRs are
driven by the net gas accretion (gas inflow minus gas outflow)
regulated by the competition between the growth of dark-matter haloes
and decrease in cosmic density \citep[\eg,][]{bouche09,dutton10}.
Therefore, measuring the galaxy SFHs constrain both the growth of
galaxy stellar mass, and the rate that gas accretion occurs.

\begin{figure}
\includegraphics[width=80mm]{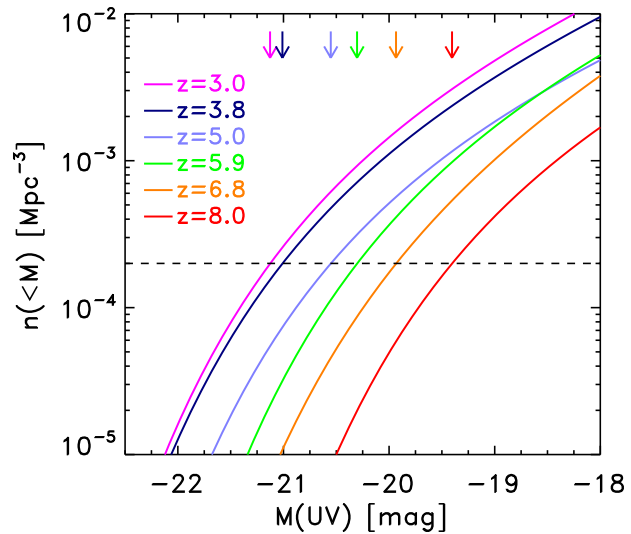}
 \caption{The comoving spatial number density of galaxies brighter
than given absolute magnitude,  $M_\mathrm{UV}$, derived by
integrating the UV luminosity functions at $z=3$ to $z=8$ using
values in table~\ref{table} from Reddy \& Steidel (2009) at $z=3.0$,
Bouwens et al.\ (2007) at $z=3.8$ to $5.9$, and Bouwens et al.\
(2010a) at $z=6.8$ and $z=8.0$.  We compared galaxies at constant number
density by identifying at each redshift the
absolute magnitude such that $n(<\!\!\muv) = 2 \times
10^{-4}$~Mpc$^{-3}$ denoted by the dashed horizontal line.  The arrows
indicate the value of $\muv$ that corresponds to this number density
for each luminosity function.   }\label{fig:lf}
\end{figure}

Here, we use observations of galaxies to measure empirically their
SFHs from $z=8$ to 3.    We compare galaxies at constant (comoving)
number density over a range of redshift because this allows us to
measure the evolution of the predecessors and descendants of galaxies
in a relative meaningful way that is not possible using other
selection methods.   Our study is motivated in part by previous
studies of galaxies at constant number density to infer
evolution \citep[see, \eg,][]{conroy09,vandokkum10}.  Studies of
galaxy properties at fixed mass or luminosity compare galaxies of very
different number density as a function of redshift,  leading to
potential biases (a variant of the well known \textit{Malmquist bias})
when comparing galaxies selected at different redshifts.    For
example, \citet{stark09} study the evolution of galaxies at constant
rest-frame UV \textit{luminosity}, finding negligible evolution in the
average stellar masses and ages of galaxies over the redshift range $4
< z < 6$.  However, the number densities of galaxies at fixed
luminosity change by factors $>$2 over this range (see
figure~\ref{fig:lf} and discussion below) implying strong evolution.
Comparing galaxy samples at constant number density has the benefit
that it allows us to follow the evolution of the progenitors and
descendants of galaxies across different redshifts (see
Appendix~\ref{section:appendixA}). 

This paper is organized as follows.  In \S~\ref{section:selection} we
discuss the properties of galaxies at constant number density at
redshifts $3 < z < 8$.  In \S~\ref{section:sfrhistory} we constrain
empirically the SFHs of galaxies at constant number density.  In
\S~\ref{section:massvsfr} we compare the joint evolution of stellar
mass and SFR for galaxies at constant number density, finding them to
be largely consistent.  In \S~\ref{section:gasmass} we use the SFR and
stellar mass evolution combined with the expected growth of galaxy
disks to predict  the evolution of galaxy gas fractions at $3 < z <
8$.   We provide discussion in support of using galaxy samples at
constant number density in Appendix~\ref{section:appendixA}, and we
discuss how models with SFRs that increase with time affect the
stellar masses of distant galaxies in
Appendix~\ref{section:appendixB}.    Throughout this paper we use a
cosmology with parameters, $H_0 = 70$~km s$^{-1}$ Mpc$^{-1}$,
$\Omega_{M,0}$=0.3, and $\Lambda_0$=0.7.  All magnitudes quoted here
are measured with respect to AB system, $m_\mathrm{AB} = 31.4 - 2.5
\log(f_\nu/1\, \mathrm{nJy})$ \citep{oke83}.

\vspace{-14pt}
\section[]{The properties of galaxies at constant number
density}\label{section:selection}

We compare galaxies at constant comoving number density, $n = 2 \times
10^{-4}$~Mpc$^{-3}$.  For redshifts $3 < z < 8$ this
corresponds approximately to the space density of $>\!\!L^\ast$
galaxies, and at $z\sim 0.1$ galaxies with this number density have
stellar masses $>\!\!1.5 \times 10^{11}$~$\msol$, approximately twice
that of the Milky Way Galaxy.   Therefore, this number density
corresponds to ``typical'' galaxies at these redshifts.

Comparing galaxies at constant number density has several advantages.
In the absence of mergers the comoving number density of galaxies is
invariant with time:  galaxies will grow in baryonic mass through gas
accretion and grow in stellar mass through star formation, but their
number density will remain unchanged.  In
principle, identifying galaxies at constant number density directly
tracks the progenitors and descendants of the galaxies at different
redshifts.  Even a non-negligible number of mergers would have only a
small effect on the progenitors and descendants of galaxies selected
at constant number density \citep[\eg,][]{vandokkum10}. In
Appendix~\ref{section:appendixA} we study how well selecting
haloes at constant number density identifies their progenitors and
descendants using simulated merger trees
\citep{spri05a}.
%
%
We found that selecting haloes at constant number density at different
redshifts identifies 60--80\% of the halo progenitors and descendants
over the entire redshift range from $3.0 < z < 7.3$, and that this
completeness fraction remains high even when selecting at constant
number density based on SFR instead of halo mass.  
%
%
 In contrast, studies of galaxies selected constant mass or
luminosity correspond to highly heterogeneous samples at different
redshifts for the reason that these quantities evolve strongly with
time.  Therefore, selecting galaxies at constant number density is a
relatively straightforward means to study the properties of galaxy
progenitors and descendants at different redshifts in a way
that is unfeasible using other galaxy selections.  

Here, we compare galaxies at constant number density based on their
UV luminosity.   
%
%
Observations of galaxy clustering show that galaxy UV
luminosity traces the halo mass tightly
\citep[\eg,][]{giav00,adel05b,kslee06,kslee09}.  Theoretically, the
relation is unclear, and predictions span a wide range for the scatter
between SFR (UV luminosity) and halo mass.   While some models
predict a tight relation, with
scatter $\lsim$0.1~dex \citep[\eg,][]{dutton10,finl10}, others predict
sizable scatter, $\sim$0.3~dex \citep[\eg,][]{delucia06,bertone07}.
To avoid an over-reliance on theoretical expectations, we make here
the simple assumption that there is a one-to-one correspondence
between the number density of galaxies at fixed halo mass and galaxies
at fixed UV luminosity, which is supported by the observational
evidence. 
%
%
%
However, in Appendix~\ref{section:appendixA} we study how scatter
in the relationship between UV luminosity and dark-matter halo mass
impacts the completeness in the descendants and progenitors of
galaxies selected at constant number density.     To summarize, we
find that increasing the scatter in this relation reduces the completeness in
the number of halo descendants identified at lower redshift.  However,
even in models where the scatter is highest, the completeness is
$>$40\% over $3 < z < 7.3$, implying that the method tracks the
descendants with high fidelity.  Furthermore, the majority of the
descendants of galaxies that are ``missed'' at lower redshifts have
halo masses within 0.3 dex of the selected galaxies.  
%
%
%
We find some empirical evidence that galaxies have average SFHs
that differ only by a scale factor related to total mass
(\S~\ref{section:sfrhistory}), consistent with expectations from
simulations \citet{finl10}.  Therefore, the descendants that are
missed likely have similar SFHs.  Even though the descendants of
galaxies at selected constant number density are incomplete, this
method allows us to study the evolution of \textit{averaged} galaxy
populations over this redshift range. 

\subsection{Selecting Samples}

\begin{table*}
 \begin{minipage}{160mm}
 \centering
  \caption{Properties of galaxies with comoving number density $n=2\times 10^{-4}$ Mpc$^{-3}$.}\label{table}
  \begin{tabular}{@{}lccccccccccc@{}}
  \hline
   $z$~\footnote{Median redshift of each sample.}  &
   $M^\ast$~\footnote{Characteristic magnitudes (masses) in units of magnitude (solar masses), faint end slope,  and space density at the characteristic magnitude (mass) in units of
     $10^{-3}$ Mpc$^{-3}$ mag$^{-1}$ ($10^{-3}$ Mpc$^{-3}$ dex$^{-1}$). }  & 
\addtocounter{mpfootnote}{-1}
   $\alpha$~\mpfootnotemark &
\addtocounter{mpfootnote}{-1}
     $\phi^\ast$~\mpfootnotemark & 
     ref.\footnote{References for luminosity or mass function parameters: \citet[RS09]{reddy09}, \citet[B07]{bouw07},
\citet[B10]{bouw10d}, \citet[M09]{marc09}.}  & 
 $A_\mathrm{UV}$~\footnote{Average attenuation at 1600~\AA\ at \muv\ in magnitudes.} & 
ref.\footnote{References for the attenuation at 1600~\AA:  \citet[B07]{bouw07},
  \citet[F10]{fink10}.} & 
  \muv~\footnote{Absolute magnitude of galaxies at comoving number density
     $n(<\!\!\muv) = 2 \times 10^{-4}$~Mpc$^{-3}$.} &
$\log \Psi$~\footnote{Extinction corrected SFR  in units of \msol\
  yr$^{-1}$ derived from \muv\  and $A_{UV}$.} & 
$\log M_\ast$\footnote{For UV-selected
  galaxies, these are the mean stellar
  mass in units of \msol\ for galaxies with \muv.  For stellar-mass-selected galaxies,
  these are the stellar masses in units of \msol\ for galaxies at comoving number
  density $n(>\!\!\mstar) = 2 \times 10^{-4}$ Mpc$^{-3}$.} & 
ref.\footnote{References for stellar masses for the UV-selected galaxies: 
\citet[R06]{reddy06a},   \citet[Sh05]{shap05}, \citet[St09]{stark09}, \citet[F10]{fink10}.  All stellar masses have been converted to a Chabrier IMF.} & 
Original IMF\footnote{Original IMF for stellar masses.  Stellar masses
  using a \citet{salp55} or \citet{kroupa01} IMF were converted to a
  Chabrier IMF by multiplying by a factor of 0.54 and 1.13, respectively. }
\\
 \hline
\multicolumn{11}{c}{Derived from the UV luminosity functions} \\
2.3 & -20.70 &  -1.73& 2.8\phantom{00} &  RS09 &
1.71 &  B09 & -21.06   & 1.64   &  9.9  & R06,Sh05 & Salpeter\\
3.1 &  -20.97&  -1.73&  1.7\phantom{00} & RS09 & 1.77 & B09 & -21.1   &
1.69   & 10.0  & R06,Sh05 & Salpeter\\
3.8 &  -21.06&  -1.76&  1.1\phantom{00} & B07  & 1.67 &  B09 & -21.0   &
1.60   &  9.6   & St09 & Salpeter\\
5.0 &  -20.69&  -1.69&  0.90\phantom{0} & B07 & 0.71 & B09 & -20.6    &
1.03   &  9.3   & St09 & Salpeter\\
5.9 &  -20.29&  -1.77&  1.2\phantom{00} & B07 & 0.06 & B09 & -20.3    &
0.67   &  9.0   & St09 & Salpeter\\
6.8 &  -20.11&  -1.94&  0.90\phantom{0} & B10 & 0.0 & F10 & -19.9
& 0.50   &  8.7   & St09 & Salpeter \\
8.0  &  -20.28& -2.00 &  0.38\phantom{0} & B10 & 0.0 & F10  & -19.4
& 0.29   &  8.6  & F10 & Salpeter \\ \hline
\multicolumn{11}{c}{Derived from the Stellar mass functions} \\
0.1 &  10.96 &  -1.18 & 3.1\phantom{00} & M09 & \ldots & \ldots & \ldots &
\ldots & 11.2  & \ldots  & Kroupa\\
1.5 &  10.91&  -0.99&  1.0\phantom{00}& M09 & \ldots & \ldots & \ldots &
\ldots & 11.0  & \ldots & Kroupa\\
2.5 &  10.96&  -1.01&  0.40\phantom{0}& M09 & \ldots  & \ldots & \ldots &
\ldots & 10.7  & \ldots & Kroupa\\
3.5 &  11.38 & -1.39 &  0.053& M09 & \ldots & \ldots & \ldots & \ldots &
10.3  & \ldots & Kroupa \\
\hline
\end{tabular}
\vspace{-12pt}
\end{minipage}
\end{table*}

Many studies have measured UV luminosity functions of galaxies from $z
\sim 2$ to $z\gsim 8$
\citep[\eg,][]{bouw04,bouw06,bouw10c,bouw10d,reddy09,oesch10b,mclure10}.
We integrated the UV luminosity functions to derive the number density
of galaxies brighter than a given absolute magnitude at each redshift,
\begin{equation}
n(<\!\!M) = \int^{M}_{-\infinity}\,\, \phi(M)\, \mathrm{d}M
\end{equation} 
where $\phi(M)$ is the luminosity function in units of number
Mpc$^{-3}$ mag$^{-1}$.  Figure~\ref{fig:lf} shows the integrated
luminosity functions at each redshift.      For the samples in this
study we measured  the evolving UV absolute magnitude, $\muv(z)$,
corresponding to the constant number density $n(<\!\!\muv) = 2\times
10^{-4}$~Mpc$^{-3}$ at each redshift.  These values are given in
table~\ref{table} and indicated in figure~\ref{fig:lf}.

The UV magnitude at constant number density $\muv(z)$ brightens with
decreasing redshift over this redshift range \citep[similar to the
evolution in $M^\ast_\mathrm{UV}$, see][]{bouw10a}.   The
values of $\muv(z)$ are based on the integral of the luminosity
functions, and they are more robust than the individual
luminosity function parameters.
%
%
While there is non-negligible uncertainty on the parameters of the
UV luminosity functions in the literature,  this has little effect on the integral of
the luminosity function.   Using the $z=8$ UV luminosity function from
\citet{mclure10} gives an absolute magnitude $\muv(z)$ within 0.1 mag
of the value  in table~\ref{table}.    

Figure~\ref{fig:lf} illustrates an importance of using galaxy samples
on the basis of constant number density rather than constant
luminosity.   Objects at constant luminosity (or stellar mass)
correspond to very different number densities at different redshifts.
The evolution in the UV luminosity functions corresponds to a change
in number density of about a factor of 5 from $z=7$ to $z=3$ at fixed
absolute magnitude, $M_\mathrm{UV} = -20$~mag.  Galaxies selected at
constant luminosity at different redshifts will result in samples with
very different evolutionary paths compared to galaxies with constant
number density \citep[\eg,][]{conroy09}. 

\subsection{Extinction and star formation
  rates}\label{section:extinction_sfr} 

While the rest-frame UV light traces the instantaneous SFR of massive stars,
it is subject to dust attenuation.   Recent evidence shows that the
mean dust attenuation decreases both with decreasing galaxy luminosity
and increasing redshift \citep{reddy08,reddy10,bouw10b,fink10}.    We
quantify the amount of extinction for galaxies with $\muv(z)$ using
published values for the dust attenuation at rest-frame 1600~\AA,
$A_\mathrm{UV}$, for star-forming galaxies as a function of
luminosity and redshift, listed in table~\ref{table}.   As there is
little evidence for dust attenuation at $z\gsim 6$
\citep{bouw10b,bouw10a,fink10} we assign $A_\mathrm{UV}=0$~mag for
these galaxies. 

We combine the UV absolute magnitudes and dust attenuation to derive
dust-corrected SFRs, $\Psi(z)$, 
\begin{eqnarray} \Psi(z)\!\!&\!\!=\!\!&\!\!10^{0.4[A_{UV}
- 48.6 - M_\mathrm{UV}(z)-\mu(z)]} \nonumber \\
\!\!&\!\!\!\!&\!\!\times4\pi d_L(z)^2 \times 7.8 \times
10^{-29}\,\,\msol\,\mathrm{yr}^{-1}.
\end{eqnarray}
where $\mu(z)$ is the distance modulus, and $d_L(z)$ is the luminosity
distance in units of cm. The numerical factor is from \citet{kenn98}
converted to a \citet{chab03} IMF.\footnote{Throughout we derive SFRs
and stellar masses assuming the IMF of \citet{chab03}.  The stellar
mass and SFR of a stellar population with a Chabrier IMF are factors
of 1.8 lower than for the IMF of \citet{salp55}, and factors of 1.13
higher than for IMF of \citet{kroupa01}.  We note that calibration
between the SFR and UV luminosity depends on the assumed SFH.  We adopt
the relation above, which assumes a constant SFH \citep{kenn98}.  The
SFR calibration for the UV luminosity for the SFH we derive below
(equation~\ref{eqn:sfrvz}) is within 30\% for the stellar population
ages considered here. } 
 
There are several sources of uncertainty on the derived SFRs.   The
scatter in the dust extinction at $\muv$ is non-negligible for
galaxies at the lower end of our redshift range where the luminosity
functions are well measured,  producing uncertainties of a factor of
order 2.  There is little dust attenuation at the high end of our
redshift range, but the luminosity function is less-well constrained,
leading to uncertainties of a  factor of order 2.  One may expect
these uncertainties to be lower when averaged over the galaxy
population at each redshift.  Lastly, the uncertainty in the UV
luminosity--SFR calibration adds an additional 0.1~dex to the error
budget (see footnote 2).  Combining these, we apply a scatter of 0.3
dex to all SFR values for galaxies at $3 < z < 8$ in this study. 

\subsection{Stellar masses}\label{section:masses}

Various studies have derived stellar masses, $M_\ast$,  for
high-redshift galaxies by modeling the
rest-frame UV-to-near-IR data
\citep[\eg,][]{shap05,papo05,papo06a,reddy06a,stark09,fink10,labbe10a}.
We use these results to derive the median stellar mass for galaxies at
each redshift brighter than $\muv(z)$.  We present these stellar
masses in table~\ref{table}.  Both the scatter in the stellar mass
distribution at \muv\ and systematics in the stellar mass estimation
contribute to the error budget \citep[see
e.g.,][]{papo06a,stark09,gonz10b}.  At $z \lsim 6$ the scatter in the
SFR--stellar mass relation dominates the random errors, and we adopt
the scatter  $d\log(M^\ast) = 0.5$~dex \citep{gonz10b}.    
%
%
%
At $z\gsim 6$, random errors from the
stellar-population modeling are comparable to the scatter in the
SFR--stellar mass relation \citep{fink10,gonz10b}.  At  $z\sim 7$ and
$z\sim 8$ we adopt a scatter of  $d\log({M^\ast}) = 0.5$~dex,
\citep{gonz10b}.  However, at $z=7$ the 68\% the lower bound on the confidence
range  extends to a factor of order 5 and at $z\sim 8$ it extends to a
factor of order 10 \citep{fink10}.  We adopt these values for the
scatter here.
%

We also derived the stellar mass for galaxies 
with $n(>\!\!\mstar) = 2 \times 10^{-4}$~Mpc$^{-3}$ at $0 < z < 4$ by
integrating the stellar mass functions of \citet{marc09} in the same
way as for the luminosity functions above.  Table~\ref{table} lists
these stellar masses, $\mstar(z)$, derived from the mass functions,
and converted to a Chabrier IMF.   At $z < 3$ the parameters of
the stellar mass functions are well constrained for the stellar masses
of interest here, and the formal uncertainty on the stellar mass at
fixed number density derived by integrating the luminosity function is
small, $<$0.1 dex.  At $3 < z < 4$ the stellar mass function is less
well constrained, and we find that the uncertainty on the mass
function corresponds to uncertainties on the stellar mass at
$n(>\!\!M^\ast) = 2 \times 10^{-4}$ Mpc$^{-3}$ of $\log (\mstar /
\msol) = 10.2^{+0.2}_{-0.7}$.  In addition, we adopt a minimum error bar of
0.3~dex \citep[see, \eg,][]{marc09} on the stellar masses at fixed
number density calculated by integrating the stellar mass functions. 

The stellar masses derived at constant number density from the mass
functions provide an important comparison sample for
the UV-selected galaxies.   At $z\lsim 3$, observations suggest that
there exist both infrared luminous and UV-luminous star-forming
galaxies \citep[\eg,][]{papo06a}, which make a substantial
contribution to the cosmic SFR history
\citep[\eg,][]{reddy08,bouw10b}.  There also exist at $z\lsim 3$
galaxies with suppressed or absent star-formation
\citep[\eg,][]{kriek06,vandokkum09b,papo10a}, although such systems
are much rarer or absent at $z\gsim 3.5$ \citep{bram07,stutz08}.   The UV
luminosity functions at $z\lsim 3$ may be incomplete for these types
of galaxies.  The stellar masses at constant number density derived
from the \citet{marc09} mass functions provide a separate measure of
the evolution of the stellar mass growth in galaxies selected by means
other than the UV luminosity.

\section[]{THE STAR FORMATION HISTORIES OF DISTANT
GALAXIES}\label{section:sfrhistory}

\begin{figure}
\includegraphics[width=84mm]{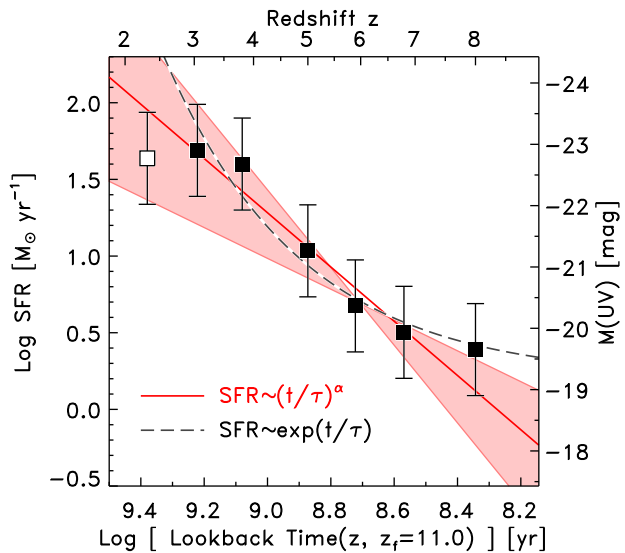}
 \caption{The SFR of galaxies at constant number density $n = 2 \times
10^{-4}$~Mpc$^{-3}$ as a function of lookback time measured from
$z=11$ to the observed redshift, $z$.  The SFRs are derived from the
UV luminosity and corrected for dust extinction as discussed in
\S~\ref{section:extinction_sfr}.  At this number density the  SFR
increases with time (decreasing redshift) from $z=8$ to $z=3$, with
data points indicated as filled symbols. The red solid lines and
shaded region show fits $\Psi(t) \sim t^{\alpha}$ to the data from $3
< z < 8$ with parameters and 68\% confidence range given in the text.
The power-law fit formally has a better goodness-of-fit than the
black dashed line, which shows a model where the SFR increases
exponentially in time, $\Psi(t) \sim \exp(t/\tau)$.  The SFR
unambiguously increases smoothly with time from $z=8$ to $z=3$.
Galaxies at $z < 3$ appear to depart from this evolution, and we
indicate the datum at $z=2.3$ with an open symbol.  }\label{fig:zvsfr}
\end{figure}
   
Figure~\ref{fig:zvsfr} shows the evolution of the dust-corrected SFR
averaged over galaxy populations with $n=2\times 10^{-4}$ as a
function of lookback time, calculated from a formation redshift
$z_f=11$  to  the observed redshift,\footnote{The choice of formation
redshift $z_f=11$ is arbitrary and does not affect our conclusions.
Changing the formation redshift adds only a constant offset in time to
equation \ref{eqn:sfrvz}. } using the data in table~\ref{table}.
There is an unambiguous smooth increase in the SFR  from $z=8$ to 3.
The relationship is consistent with a power law, described by
\begin{equation}\label{eqn:sfrvz}
\frac{\Psi(z)}{1\, \msol\ \mathrm{yr}^{-1}} = (t/\tau)^{\alpha}
\end{equation} 
with $\alpha = 1.7\pm 0.2$ and $\tau=180\pm 40$~Myr fitted over the
data from $z=8$ to 3.  We adopt here the power-law fit as it formally
has a better goodness-of-fit compared to other simple functions,
although it is not a unique representation of the data.  Fitting a
model to the data where the SFR increases exponentially in time,
$\Psi(z) = \Psi_0 \exp(t/\tau)$, gives a minimum $\chi^2 = 1.1$ for 4
degrees of freedom (for $\tau=420$~Myr and $\Psi_0=1.4$~\msol\
yr$^{-1}$), compared to $\chi^2 = 0.66$ for 4 degrees of freedom for
the power-law fit above.    However, statistically these fits are
indistinguishable statistically.

Here, we concentrate on the evolution of galaxy properties at constant
number density $n = 2 \times 10^{-4}$ because this describes the
evolution of ``typical'' galaxies (\S~2),  and this is where the UV
data are most complete (see references in table~\ref{table}).
However, we also studied how the SFH varies with galaxy number
density, ranging from $n=1 \times 10^{-3}$~Mpc$^{-3}$ (corresponding to
fainter galaxies with $\sim$$L^\ast + 1$~mag) to $n=4\times
10^{-5}$~Mpc$^{-3}$ (corresponding to brighter galaxies with $\sim$$L^\ast -
0.5$~mag).    We observe a modest change in the slope of
equation~\ref{eqn:sfrvz} for these samples, from $\alpha=1.5$ at
$n=1\times 10^{-3}$~Mpc$^{-3}$ to $\alpha=1.9$ at $n=4\times
10^{-5}$~Mpc$^{-3}$, but these are consistent within the measured
uncertainty, $\delta\alpha$=0.2.  However, the normalization
($\tau^{-\alpha}$) changes relative to that for $n=2 \times
10^{-4}$~Mpc$^{-3}$ by  a factor of $1/22$ at $n=1\times
10^{-3}$~Mpc$^{-3}$ to a factor of 12 at $n=4\times
10^{-5}$~Mpc$^{-3}$.  Therefore, over the range of number densities
considered here ($n = 4 \times 10^{-5} - 1 \times 10^{-3}$
Mpc$^{-3}$) galaxies experience SFRs that increase as a power-law in
time with a slope, $\alpha$, that is approximately constant.  The SFR scales only by a
multiplicative factor related to number density \citep[or mass,
similar to the simulations of][]{finl10}, where galaxies with higher
number density (lower UV luminosity) experience slower evolution
\citep[\eg,][]{bouw10d}. 
%

At $z < 3$ the SFRs derived for UV-selected galaxies appear to depart
from the SFH in equation~\ref{eqn:sfrvz} (figure~\ref{fig:zvsfr}).  At
$z \sim 2-3$ galaxies with UV luminosities near $L^\ast$ show a
significant heterogeneity in their dust properties, and galaxies at
fixed UV luminosity include both intrinsically less luminous galaxies
with less dust extinction and intrinsically more luminous galaxies
whose UV emission is obscured by dust \citep{reddy10}, and so we may
expect the departure from the derived SFH.   However, our derived SFH
is consistent with the space density of $K$-selected ultraluminous IR
galaxies (ULIRGs) at $1.5 < z < 2.5$, which have $n=2\times
10^{-4}$~Mpc$^{-3}$ and $\langle\Psi\rangle$ = 120 \msol\ yr$^{-1}$
\citep{daddi07a}.  Therefore, while UV-selected samples seem
incomplete for star-forming galaxies at redshifts $z\lsim 3$, the
derived SFH remains consistent when we include observations of
obscured galaxies.

Therefore, galaxies with this constant number density have average
SFRs that increase with time for redshifts $3 < z < 8$.   This is
similar to expectations from theory
\citep[\eg,][]{robe04,delucia06,finl06,finl10,brooks09}, which predict
rising SFHs over this redshift range.     We stress that the SFHs we
derived here describe the \textit{cosmologically average} evolution of
the galaxy population at these redshifts, and individual galaxies
likely experience unique SFHs, including stochastic bursts and periods
of reduced star-formation.    In particular, at later times, $z \lsim
3$, the SFHs of galaxies seemingly plateau, and at later times
galaxies must experience SFHs that decline with time in order to
explain the existence of lower redshift galaxies with evolved stellar
populations evidenced by the existence of strong 4000~\AA/Balmer
breaks by $z\sim 2$ \citep[\eg,][]{kriek06}.  Nevertheless, as  we
discuss below (\S~\ref{section:massvsfr}), the \textit{average}
smoothly rising SFHs for the galaxy population at early times
reproduce the stellar mass growth in galaxies.  We therefore,
conclude that galaxies with this constant number density experience
rising SFHs during the epoch demarcated by $3 < z < 8$.

\begin{figure}
\includegraphics[width=94mm]{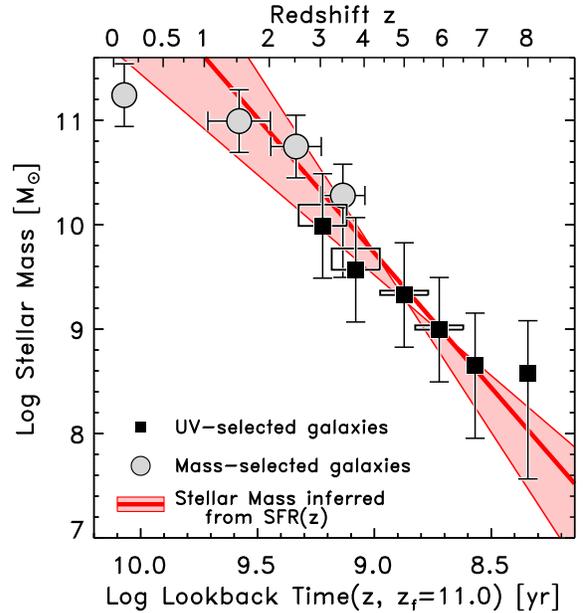}
 \caption{ The stellar mass of galaxies as a function of lookback time
for galaxies with constant number density. Filled squares show the
stellar masses for galaxies with UV luminosities such that $n(<\muv) =
2\times 10^{-4}$~Mpc$^{-3}$ from $z=8$ to 3.  The heavy red line and
shaded region shows the possible range of stellar mass evolution
expected from SFH measured in figure~\ref{fig:zvsfr}.  This line is
not a fit, as we have made no adjustments to the normalization or
slope of this line.    The stellar mass growth expected from the
measured SFH agrees with the data for UV-selected galaxies at $3 < z
< 8$.   The measured masses for UV samples at $z < 4$ are slightly
below the expected evolution for possible reasons discussed in the
text. 
%
%
The filled
circles show the evolving stellar masses for galaxies with stellar
mass such that $n(>\!\!\mstar) = 2 \times 10^{-4}$~Mpc$^{-3}$
\citep{marc09}. These follow the expected stellar masses from the SFH
for redshifts $z\gsim 1$.  Therefore, the SFH and stellar mass growth are
consistent over this redshift range. }\label{fig:zvmass}
\end{figure}

\section{THE EVOLVING SFR-STELLAR MASS RELATION}\label{section:massvsfr}

The measured SFH provides a direct prediction for the stellar mass
evolution for galaxies at this constant number density.    For a SFR
that evolves as $\Psi(t) \sim t^\alpha$ (equation~\ref{eqn:sfrvz})
the stellar mass evolves roughly as $M_\ast(t) \sim \int\, \Psi(t)\,
dt \sim t^{(\alpha+1)}$.  However, one must account for stellar-mass
loss due to baryonic gas returned to the ISM of a galaxy from stellar
evolution.  We therefore computed the expected stellar mass evolution
using the SFH from equation~\ref{eqn:sfrvz} with the 2007 version of
the \citet{bruz03} stellar-synthesis model.  We use an input stellar
population with $Z=0.2$~\zsol\ metallicity and a Chabrier IMF with no
chemical evolution.   

Figure~\ref{fig:zvmass} shows the evolving stellar mass, $\mstar$, as
a function of lookback time for galaxies with $n=2\times
10^{-4}$~Mpc$^{-3}$.   The measured masses of the UV-selected galaxies
at this number density are shown as black squares.  The heavy red line
and shaded region in figure~\ref{fig:zvmass} shows the
\textit{expected} increase in the stellar mass from the range of SFHs
in figure~\ref{fig:zvsfr}.      It is important to note that the red
curves in figure~\ref{fig:zvmass} are not fits to the data, but rather
predictions from the derived SFH. 
%

The stellar mass evolution of the UV-selected galaxies at $3 < z < 8$
agree well with this expected growth from the measured SFR history.
At $z \sim 8$ the stellar mass for UV-selected galaxies with this
number density lie above the expected value, although still within the
scatter.  

At $z \lsim 4$ there is slight evidence that stellar masses for the UV
samples fall  below the expected trend by about 0.3~dex (although
again within the scatter).  This could be symptomatic of older stellar
populations with high mass-to-light ratios  which are plausible to the
limits of the data \citep[\eg,][]{papo01,shap05}.  Alternatively, the
stellar masses derived for the UV samples assumed SFHs that are either
constant or decline with time, rather than increase with time.  As
discussed in Appendix~\ref{section:appendixB}, stellar masses from
rising SFHs are larger by $\approx$0.2~dex for galaxies at $z\sim 3-4$
compared to models with SFRs that are constant or decline with time
\citep[see also][]{sklee10}.  In contrast the stellar masses from
models with rising SFRs are only $<$0.1 dex higher at $z\sim 5-6$.
These shifts are indicated by the open boxes in
figure~\ref{fig:zvmass}, and they would account for most of the
observed offset.    Furthermore, this offset may be also indicative of
the fact that at $z\sim 2-3$ galaxies with UV luminosities near
$L^\ast$ show a significant heterogeneity in their dust properties
\citep[see \S~\ref{section:sfrhistory}, and][]{reddy10}, or as a
result of the scatter in the UV-luminosity--dark matter relation (see
Appendix~\ref{section:appendixA}). Lastly, by $z\lsim 3$ there exist
massive galaxies with ``suppressed'' SFRs \citep[\eg,][]{kriek06},
which could be missed by the UV selection.  These effects may bias the
derived stellar masses for these galaxies to lower values.   In
contrast, these effects are less significant at higher redshifts
$z\gsim 4$ because there has been less time for them to  manifest. 

Because of potential biases in the UV-luminosity-selected sample,
figure~\ref{fig:zvmass} also shows the stellar masses for galaxies at
$z < 4$ with constant number density, $n(>\!\!\mstar) = 2 \times
10^{-4}$~Mpc$^{-3}$, derived from the mass functions of
\citet{marc09}.   The selection based on \textit{stellar mass}
mitigates most of the biases for the UV galaxies.   Remarkably, the
evolution of stellar mass for galaxies at constant number density
agrees well with the stellar mass growth expected from the SFH for
$z\gsim 2$.  
This supports our conclusion that the cosmologically averaged
SFHs of distant galaxies are smoothing rising from $z=8$ to 3 because
this matches both the increased UV luminosities and stellar masses of
galaxies with this number density.  

The models with rising SFHs contrast assumptions that distant galaxies
experience SFRs that are either constant or decreasing with time
\citep[\eg,][]{papo01,papo05,papo06a,shap01,forster04b,shap05,reddy06a,forster09,stark09,fink10,labbe10a,labbe10b}.
Several other studies have suggested SFHs that increase with time
based on the evolution in the SFR-stellar mass relation at high
redshifts ($z\lsim 6$) \citep{renz09,stark09,mara10} and also based on
theoretical motivation \citep{finl07,finl10}.  Our analysis
demonstrates unambiguously that the cosmologically averaged SFRs for
typical galaxies increase smoothly with time (decreasing redshift).
Individual galaxies likely deviate from this ``smoothly'' increasing
SFRs because they have formation histories that involve stochastic
events with discrete changes in the instantaneous SFR
\citep[\eg,][]{baugh05,finl06,finl07,some08}, and they may experience
SFHs that deviate from the average history derived here.  The SFHs
that we derive here corresponds to a cosmologically averaged evolution
that accounts for the observed evolution of the galaxy populations.

The self-consistency between the rising SFHs and the stellar mass
growth constrains jointly (degenerately) both the star formation duty
cycle and the slope of the high-mass end of the IMF.  Observations of
clustering measurements and the correlation and scatter of the
SFR-stellar mass relation of high--redshift galaxies imply a high
star-formation duty cycle \citep[\eg,][]{daddi07a,labbe10a,kslee10}.
If the duty cycle is near unity for galaxies at $3 < z < 8$,  then the
consistency between the rising SFHs and stellar mass growth here is
evidence that the high-mass end of the IMF in these galaxies must have
a slope consistent with that of \citet{chab03} and \citet{salp55} when
averaged over galaxy populations.  Otherwise, there would be a larger
offset in the SFR-stellar mass relation, which is not evident in the
data.  

If the star formation duty cycle is lower as suggested by clustering
analyses of (predominantly) sub-$L^\ast$ sources \citep[\eg,][although
see Finlator et al.\ 2010]{kslee09} and by some models of galaxy
formation \citep[\eg,][]{baugh05}, then our integrated SFH requires an
IMF weighted toward higher mass stars in order to match the measured
stellar mass growth.    For example, a duty cycle of 50\% would
require a change of $\approx 0.2$ to the slope of the IMF to produce a
higher ratio of the number of O-type stars to the number of G-type
stars compared to that produced by a Salpeter-like IMF.   This is
similar to the findings of \citet{baldry03} based on SFHs derived from
the UV to IR luminosity densities of low-redshift galaxies.
Such top-heavy IMFs have been also proposed to explain the joint cosmic
evolution of the stellar mass and SFR densities
\citep[\eg,][]{baugh05,dave08,lacey08}, although these are more
extreme IMFs than the change described above.  Lastly, our results are
unable to constrain the low-mass end of the IMF as any change in the
shape of lower mass limit simply scale both the SFR and stellar mass
in approximately the same way, shifting both the datapoints and the
curve in figure~\ref{fig:zvmass} by approximately the same amount.

\section{IMPLICATIONS FOR THE GAS MASS AND GAS
ACCRETION RATE}\label{section:gasmass}

The evolution in figures~\ref{fig:zvsfr} and \ref{fig:zvmass} leads to
the conclusion that galaxies increase \textit{both} their SFRs and
stellar masses over time.    This simply is not possible if galaxies
form with a fixed initial cold baryonic gas mass.  In this case the
SFR would decline with time as the gas mass is converted to stars.    

In this section we use our measured evolution in the SFRs and stellar
masses of galaxies to predict the evolution of the gas mass  in
galaxies with with $n=2 \times 10^{-4}$ Mpc$^{-3}$ at high redshifts.
Throughout this section we assume that the local gas-surface density
($\Sigma_\mathrm{gas}$) - SFR surface density ($\Sigma_\mathrm{SFR}$)
relation \citep{schmidt59,kenn98b} applies to high--redshift galaxies,
\begin{equation}\label{eqn:schmidt} \frac{\Sigma_\mathrm{SFR}}{\msol\,
\mathrm{yr}^{-1} \mathrm{kpc}^{-2}} = A \left(
\frac{\Sigma_\mathrm{gas}}{\msol\,\mathrm{pc}^{-2}} \right)^N,
\end{equation} 
where $A=1.7\times 10^{-4}$ and $N=1.4$ have their normal values
\citep[though we use the calibration for a Chabrier-like IMF, see][see
also Dutton et al.\ 2010]{some08}.    The assumption that
star--formation in high--redshift galaxies follows this relation
remains untested directly at these redshifts, and such tests
will require direct measurements of the HI gas, which is currently
infeasible.    However, recent studies for example by \citet{daddi10b}
suggest that the star-forming efficiency in normal star-forming disk
galaxies is consistent with this assumption, though it may not apply
in extreme classes of star-forming galaxies.  As the ``typical''
galaxies studied have SFRs and properties consistent with ``normal''
star-forming disk galaxies at $z\sim 2$, our assumption is
reasonable. 

We multiply equation~\ref{eqn:schmidt} by the areal size of the
star-forming region in galaxies, approximated by $a = \pi r_{1/2}^2$,
where $r_{1/2}$ is the familiar half-light radius, which relates the
gas-mass to the SFR, 
\begin{equation}\label{eqn:sfrgas} \frac{\Psi(z)}{1\, \msol\,
\mathrm{yr}^{-1}} = C  \left[\frac{r_{1/2}(z)}{1\,
\mathrm{kpc}}\right]^{-0.8} \left[ \frac{M_\mathrm{gas}(z)}{10^9\,
\msol}\right]^{1.4},
\end{equation}
where the constant of proportionality is $C=1.7$.  While the statistical
uncertainty on equation~\ref{eqn:sfrgas} is 0.12~dex  \citep{kenn98b},
the true uncertainty may be larger at high redshifts where these relations have not
been measured directly. 

The rest--frame UV size evolution of high-redshift galaxies has been
characterised by \citet{ferg04}, 
%
%
who showed that the size
evolution for star-forming galaxies from $z\sim 5$ to $z\sim 2$ is
consistent with the growth of galaxy disks that scale with the size of
their dark matter haloes \citep{mo98}, $r(z) \propto H(z)^{-1}$, where
$H(z)$ is the Hubble constant at redshift $z$ \citep[consistent
with][]{oesch10a}.      We normalize this relation to produce the
characteristic half-light size at $z=4$ from Ferguson et al.,
$r(z\!\!=\!\!4) = 1.7$~kpc with a scatter of approximately 0.6~kpc.
The galaxy sizes then evolve as
\begin{equation}\label{eqn:size}
r(z) = 1.7\, \left[ \frac{H(z)}{H(z\!=\!4)} \right]^{-1}\, \mathrm{kpc},
\end{equation} 
where $H(z\!\!=\!\!4) = 430$~km s$^{-1}$ Mpc$^{-1}$.  The extrapolated
size evolution from this equation reproduces the sizes measured for
$z=7$ galaxies in \hst/WFC3 data, $r(z\!\!=\!\!7) = 0.85$~kpc,
consistent with the observations of \citet{oesch10a}.  

Combining equations~\ref{eqn:sfrgas} and \ref{eqn:size},  
and solving for the gas mass yields
\begin{eqnarray}\label{eqn:mgasvz}
\frac{M_\mathrm{gas}(z)}{5.9 \times 10^8\, \msol} = \left[ \frac{\Psi(t) }{1\,\msol\,\mathrm{yr}^{-1}} \right]^{5/7} \left[ \frac{r(z\!\!=\!\!4)}{1.7\, \mathrm{kpc}} \right]^{4/7}  \\ \nonumber
\times\left[ \frac{H(z)}{H(z\!\!=\!\!4)} \right]^{-4/7}. 
\end{eqnarray} The uncertainty on the gas mass includes the scatter in
the galaxy size measurements, and includes the uncertainty from the
SFR-gas mass surface density relationship (a factor of $\delta\log
M_\mathrm{gas} =0.11$~dex).    However, as discussed above, equation
\ref{eqn:mgasvz} implicitly assumes the star-formation law from the
relation given in equation~\ref{eqn:schmidt} and neglects any
uncertainty on this relation at high redshift.
%

\subsection{Gas mass and gas accretion rate}

\begin{figure}
\includegraphics[width=84mm]{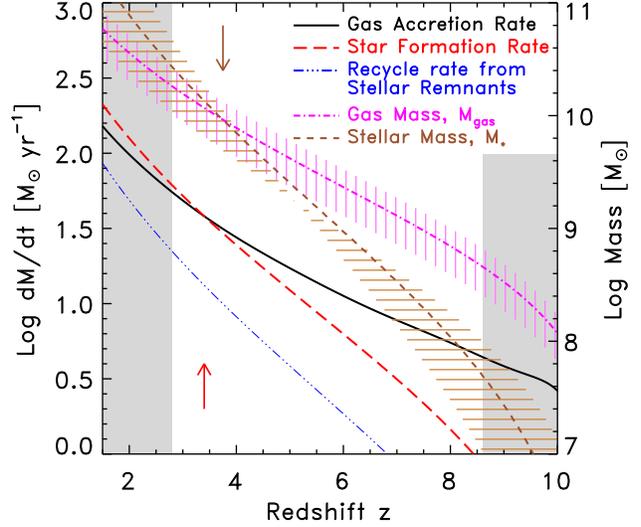}
\caption{Predicted redshift evolution of the gas accretion rate,
$\dot{M}_\mathrm{acc}$ and the gas mass, the derived stellar mass
inferred from the empirical SFH, and the empirically derived evolution
of the SFR for  galaxies with constant number density $n = 2 \times
10^{-4}$~Mpc$^{-3}$.  Here, $\dot{M}_\mathrm{acc}$ is the net gas
accretion rate and does not distinguish between the contributions of
gas inflow and outflow.  The gas mass and gas accretion rate are
discussed in the text, and assume the Kennicutt--Schmidt conversion
between gas and star-formation surface density.    Each curve shows an
evolving quantity as indicated in the legend.  Gray-shaded regions
show redshifts where there is no data, and the curves are
extrapolations.  The left abscissa shows the scale for the gas
accretion rate, the SFR, and the rate of gas mass returned from
stellar remnants.  The right abscissa shows the scale for gas mass and
stellar mass.   The downward arrow indicates the redshift where the
stellar mass equals the gas mass.  The upward arrow indicates the
redshift where the SFR equals the gas-mass accretion rate.
}\label{fig:gasacc}
\end{figure}

Figure~\ref{fig:gasacc} shows the derived redshift evolution of the
stellar mass (\mstar) inferred from the empirical SFH
(\S~\ref{section:massvsfr}),  and the predicted redshift evolution for
the gas mass (\mgas) from equation \ref{eqn:mgasvz} for galaxies with
constant number density $n=2 \times 10^{-4}$~Mpc$^{-3}$.  The stellar
mass increases with time, as discussed in \S~\ref{section:massvsfr}.
The gas mass also increases from $z=8$ to $z=3$, which
is required to fuel the increasing SFRs.  Our empirical relations
predict that from $z=10$ to $z=8$ the gas mass is an order of
magnitude larger than the stellar mass for galaxies with $n=2\times
10^{-4}$~Mpc$^{-3}$.   The gas mass continues to increase at $z < 8$.
While the stellar mass initially lags behind the gas mass, it
increases quickly in response to the SFR.  Around $z \approx 4$, the
stellar mass equals the gas mass for galaxies with this constant
number density, and extrapolating to lower redshift the stellar mass becomes the
dominant baryonic component for galaxies at this number density.

We use the derivative of the gas mass and the SFR to predict the gas
accretion rate, $\dot{M}_\mathrm{acc}$, for these galaxies.  The gas
accretion rate could be driven by either smooth accretion
\citep[\eg,][]{keres09a}, or gas delivered via significant mergers
\citep[\eg,][]{robe08,stew08}, and differences in these processes do
not affect the simple empirical relations we derive here.      Here we
do not differentiate between possible physical processes that
contribute to the net gas accretion rate but rather we define this to
be simply the net rate at which galaxies acquire their baryonic gas.
Therefore, the net gas accretion rate here includes the effects of
both gas infall and gas outflows \citep[see, \eg,][]{dave06,dutton10}.
The time-rate-of-change of the gas mass has several components,
\begin{equation}\label{eqn:dmdt}
\dot{M}_\mathrm{gas}(t) = \dot{M}_\mathrm{acc}(t) - \Psi(t) +
\dot{M}_\mathrm{rec}(t),
\end{equation}
where $\dot{M}_\mathrm{rec}$ is the amount of gas returned
(``recycled'')  to the galaxy from stellar remnants (taken from the
2007 version of the Bruzual \& Charlot 2003 models). 

Figure~\ref{fig:gasacc} shows the predicted gas accretion rate for
galaxies with number density, $n=2 \times 10^{-4}$~Mpc$^{-3}$ using
equation~\ref{eqn:dmdt} .  The figure also shows the SFR and
gas-recycling rates.     The gas accretion rate increases from $z=8$
to 3 and keeps pace with or exceeds the measured SFR.  At these
redshifts these galaxies acquire gas at least as rapidly as it can be
converted into stars -- this is the ``gas accretion epoch'' for these
galaxies, and is similar to inferences drawn from SPH simulations of
galaxy formation \citep{keres05}.    At $z\lsim 4$ the gas-accretion
rate declines, and galaxies consume their gas at a rate faster than
they acquire it.

Extrapolating the trends in figure~\ref{fig:gasacc} to lower redshifts
($z < 3$), our model shows that the stellar mass of galaxies with
$n=2\times 10^{-4}$~Mpc$^{-3}$ exceeds the available gas mass and the
SFR exceeds the gas-accretion rate.  These trends can not continue
\textit{ad infinitum} as the available gas will eventually be
consumed.  Indeed, figure~\ref{fig:zvmass} indicates that at redshifts
$z < 2-3$, the stellar masses of galaxies with $n = 2 \times
10^{-4}$~Mpc$^{-3}$ depart from the expected growth based on the
evolving SFR, and this appears consistent with the evolution of
neutral gas in absorption-line studies \citep{proc09}.      By $z\sim
2$ some galaxies with the number density considered here have low or
suppressed SFRs \citep{kriek06}, which presumably have low cold-gas
fractions.   This implies that the gas accretion rate is decreasing at
these redshifts,  consistent with  galaxy formation simulations
\citep{keres05}.   Several physical effects and feedback processes
have been proposed to prohibit gas cooling and accretion in galaxies
\citep[\eg,][]{birn03,spri05b,phopk06,phopk08,croton06,some08}, which
should apply at the mass scales of galaxies with this number density
at these redshifts. 

The semi-analytic galaxy formation models of \citet{dutton10} find
that star-forming galaxies at high redshift have SFRs in a steady
state with the total gas mass rate-of-change (gas accretion minus gas
losses due to outflows) of $\dot{M}_\mathrm{gas} \approx \Psi(t)$.
The gas accretion rate predicted by our model seems consistent with
this conclusion (although large disparities between the gas accretion rate
and SFR are not excluded).  \citet{dutton10} interpret the
relative consistency between the SFR and net gas accretion rate as a
result of increased gas densities in galaxies. Our empirical results
may support this scenario.

\subsection{Predictions for galaxy gas-mass fractions}

\begin{figure}
\includegraphics[width=84mm]{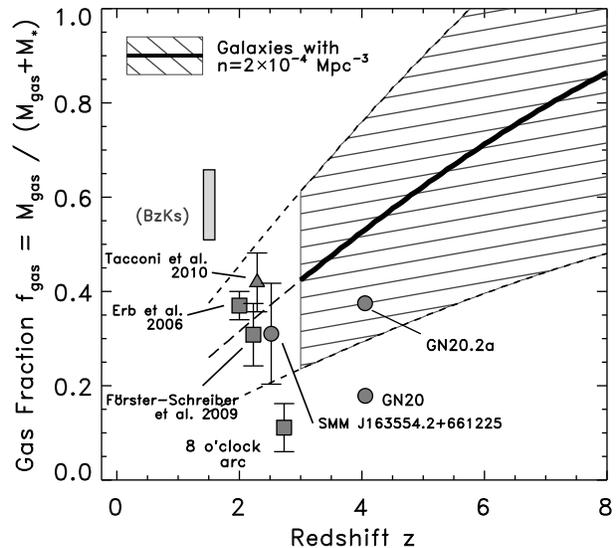}
 \caption{ Empirical constraints on the gas fraction, $f_\mathrm{gas}
= M_\mathrm{gas} \times (M_\mathrm{gas} + M_\ast)^{-1}$.  The hatched
region shows the expected range of average gas fractions for galaxies
with constant number density $n = 2 \times 10^{-4}$ for $z=8$ to
$z=3$.  The dashed lines indicate these relations extrapolated to
$z=1.5$.    Data points are labelled, and include UV-selected galaxies
\citep[filled boxes,][]{erb06b,fink09} and sub-mm-selected galaxies
\citep[filled circles,][]{kneib05,papo09,daddi09a}.  The data points
for \citet{forster09} and \citet{tacc10} show the mean value for
galaxies with $z \geq 2$ and  $M_\ast > 3\times 10^{10}$~\msol, the
approximate stellar mass at $z\approx 2-2.5$ such that $n(>\!\!\mstar)
= 2 \times 10^{-4}$ Mpc$^{-3}$.  Our empirical relation generally
accounts for the measured gas fractions of these galaxies.    There is
evidence of mass-dependent evolution.  Galaxies with higher stellar
masses have lower gas fractions  (e.g., the 8 o'clock arc, Finkelstein
et al.\ 2009; GN20, Daddi et al.\ 2009), implying faster
gas-consumption.   Likewise, lower mass galaxies have higher gas
fractions than what is predicted from the extrapolated relationship
for galaxies with $n=2\times 10^{-4}$~Mpc$^{-3}$ \citep[see also the
lower mass samples of F\"orster-Schreiber et al.\ 2009 and Tacconi et
al.\ 2010]{daddi09b} .  }\label{fig:zvfg}
\end{figure}

We determine the gas fraction as $f_\mathrm{gas} = M_\mathrm{gas}(z) \times
[M_\mathrm{gas}(z) + M_\ast(z)]^{-1}$ 
%
%
using the evolution of the stellar mass inferred from the SFH
(\S~\ref{section:massvsfr}) and the evolution of the gas mass from
equation~\ref{eqn:mgasvz}.     The hatched region in
figure~\ref{fig:zvfg} shows the predicted range of the gas fraction
from $z=8$ to $z=3$ for galaxies with constant number density $n = 2
\times 10^{-4}$~Mpc$^{-3}$. The dashed
lines show the relation extrapolated to lower redshift.  The gas mass
fraction for these galaxies decreases with decreasing redshift,
parametrised as $f_\mathrm{gas} \propto (1+z)^\Gamma$ with $\Gamma
\approx 0.9$.  This follows from the evolving stellar mass and gas
mass derived above.  At $z\approx 3-5$ the range of possible gas-mass
fractions reach 50\%, similar to our conclusions from
figure~\ref{fig:gasacc}.   

Figure~\ref{fig:zvfg} shows that the predicted gas fractions are
consistent with available observations of galaxies at $z\sim 2$ with
stellar masses such that $n(>\!\!\mstar) = 2 \times 10^{-4}$
Mpc$^{-3}$, including galaxies with gas masses derived from CO
measurements \citet{kneib05,daddi09a,papo09,tacc10} and those derived
from the SFR surface density inferred from \ha\ measurements
\citep{erb06b,fink09,forster09}.  There is evidence that the evolution
of the gas fractions is mass dependent.  Figure~\ref{fig:zvfg} also
shows gas fractions of several galaxies with masses much greater than
the galaxies with $n = 2 \times 10^{-4}$~Mpc$^{-3}$ (these are
presumably rarer systems, with lower number density).  This includes
the gravitationally lensed UV-selected, ``8 o'clock arc'' at $z=2.7$
with high stellar mass ($\gsim 10^{11}$~\msol) and low gas fraction,
which may have experienced accelerated evolution \citep{fink09}.
Similarly, the massive sub-mm-selected galaxy GN20 at $z=4$ has a low
gas-mass fraction \citep{daddi09a} \citep[see also,][]{mich10}.  These
systems show evidence of mergers, enhancing the
gas-consumption rate and/or the star-formation efficiency
\citep{daddi10b}.    Because the empirical gas fractions we derive for
galaxies with $n=2\times 10^{-4}$~Mpc$^{-3}$ agree reasonably with the
available observations, galaxies experiencing enhanced gas-consumption
rates because of significant mergers seem to be the exception, not the
norm at these redshifts. 

Galaxies with lower stellar masses than those selected with constant
number density, $n=2 \times 10^{-4}$~Mpc$^{-3}$, appear to have higher
gas-mass fractions \citep[see also the lower mass galaxies in the
samples of  F\"orster-Schreiber et al.\ 2009 and Tacconi et al.\
2010]{daddi09b}.  This is further evidence suggesting that the
gas-accretion and gas consumption time-scales dependent on galaxy
mass.

Our empirical results make very clear predictions for the
cosmologically averaged gas masses 
of galaxies with number density $n=2 \times 10^{-4}$~Mpc$^{-3}$, and also
that the gas fractions should increase as function of redshift.  While
the predicted gas masses are consistent with the limited available
data, more observations are needed.  Future observations from the
\textit{Atacama Large Millimeter Array}
(ALMA) will have the sensitivity to detect emission from the predicted gas
reservoirs in these galaxies, which will test the predictions 
on the gas masses we make here. 

\section{SUMMARY}

We use observed relations of the SFRs and stellar masses for galaxies
to constrain empirically the SFHs of  galaxies at $3 < z < 8$.  We
compare the evolution of galaxies at these redshifts at
constant comoving number density, $n(<\!\!\muv) = 2 \times
10^{-4}$~Mpc$^{-3}$.  We show that this allows us to track the
evolution of stellar mass and star formation in the direct
predecessors and descendants of these galaxies in a relatively
meaningful way that is not possible using samples selected either at
constant luminosity or constant mass.  We summarize our findings as
the following. 

The galaxies experience cosmologically averaged SFHs with SFRs that
grow with time (decreasing redshift)  at a rate best described by
$\Psi(t) \sim t^{\alpha}$ with $\alpha = 1.7 \pm 0.2$.  This excludes
model SFHs where the SFR is either constant or declines exponentially
in time.   We find evidence that galaxies over a range of number
density from $n=4\times 10^{-5}$ Mpc$^{-3}$ to $1\times 10^{-3}$
Mpc$^{-3}$ (corresponding to galaxies with luminosities
$<L_\mathrm{UV}^\ast - 0.5$~mag and $<L_\mathrm{UV}^\ast + 1$ mag,
respectively), experience SFHs with similar power-law increases with
time, but differ by a multiplicative scale factor.  These results are
similar to  expectations from theory \citep[\eg,][]{finl10}.    We
reemphasize that these SFHs correspond to the cosmologically averaged
evolution of the galaxy populations with these number densities.
Individual galaxies experience stochastic events and unique SFHs that
deviate from the average evolution derived here.

We show that the measured stellar masses for galaxies selected with
this constant number density  are consistent with the expected
stellar mass growth from the derived SFH.  If star-forming galaxies
at high redshifts have star-formation duty cycles near unity, then
this implies that the high-mass end of the stellar initial mass
function is approximately Salpeter.   For lower duty cycles, our
results would require an IMF with high-mass slope weighted toward
higher mass stars (by $\approx 0.2$ for a duty cycle of
$\approx$50\%).   Otherwise, there would be a larger offset in the
SFR-stellar mass relation, which is not evident in the data.

We interpret the relation between SFR and stellar mass as a result of
net gas accretion (gas outflows combined with gas infall from either
smooth accretion or delivered by mergers) coupled with the growth of
galaxy disks, where the SFR depends on the local gas-surface density
relation.  This leads to predictions for the galaxy gas masses and
gas accretion rates on galaxies $z=8$ to $z=3$.  The gas fraction
evolves at a rate of $f_\mathrm{gas} \propto (1 + z)^\Gamma$, with
$\Gamma \approx 0.9$, and we show these gas fractions are in
reasonable agreement with available data at $z\sim 2$.   These
relations imply that at $z > 4$ galaxies selected at this constant
number density acquire baryonic gas at least as fast as it can be converted
into stars.  This is the ``gas accretion epoch'' for these galaxies.
Interpreting this in the context of a semi-analytic model
\citep{dutton10} this suggests that the smoothly rising SFHs are a
consequence of increased gas densities in these galaxies.    At $z <
4$ the SFR in these galaxies overtakes the gas-accretion rate,
indicating a period where these galaxies consume gas faster than it is
acquired.  At redshifts $z \lsim 3$, galaxies depart from these
relations implying that gas accretion is slowed  or prevented at later
times.   
%
%

\section*{Acknowledgments}

The authors acknowledge stimulating conversations with our colleagues,
in particular the authors thank Rychard Bouwens, James Bullock,
Emanuele Daddi, Romeel Dav\'e, Darren DePoy, Kristian Finlator, Carlos
Frenk, Kyung-Soo Lee, Umberto Maio,  Michal {Micha{\l}owski}, Jason
Prochaska, Ryan Quadri, Kim-Vy Tran, and Risa Weschler for highly
interesting discussions, comments, and valuable feedback.  The authors
also wish to thank the anonymous referee whose comments and
suggestions significantly improved both the quality and clarity of
this work.  Support for this work was provided to the authors by the
George P. and Cynthia Woods Mitchell Institute for Fundamental Physics
and Astronomy.   We acknowledge our appreciation to the Virgo
Consortium for making the Millennium simulation available online.
The Millennium Simulation databases used in this paper and the web
application providing online access to them were constructed as part
of the activities of the German Astrophysical Virtual Observatory.

\bibliography{apj-jour,alpharefs}{}

\appendix

\section{ON GALAXIES AT CONSTANT NUMBER DENSITY}\label{section:appendixA}

In this paper we have studied the evolution of galaxies from $3 < z <
8$ at fixed number density.  This has the advantage that in principle
it traces the evolution of the progenitors and descendants of galaxies
as a function of redshift in a way that is not possible using other
methods such as selection at constant luminosity or mass, quantities
which themselves evolve with time.   It is also relatively
straightforward to compare results from observational galaxy surveys
to expectations from theory.  \citet{vandokkum10} applied a similar
selection to study the size evolution of massive galaxies at $z \lsim
2$. They quantified the effects of mergers on their selection using
Monte Carlo simulations of the mass function, and concluded that
selection at constant number density produces homogeneous samples at
different redshifts with low contamination unless haloes grow only via
one-to-one mergers.  Based on theoretical arguements they
argued that evolution driven by major mergers is highly unlikely
\citep[\eg,][]{naab07,guo08}.  

\begin{figure}
\includegraphics[width=84mm]{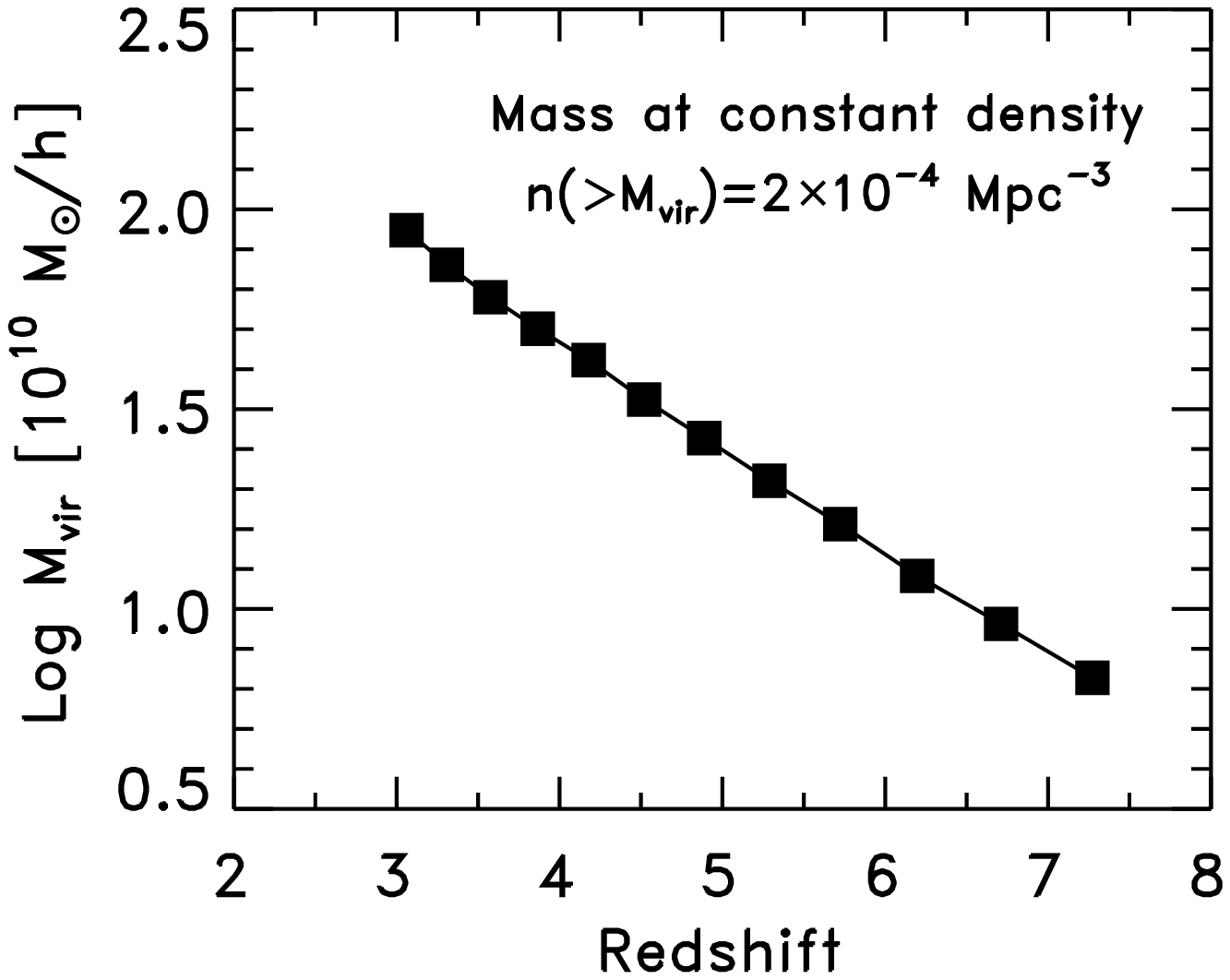}
 \caption{ The evolution of halo mass at constant
number density from the Millennium simulation \citep{spri05a}.
The datapoints show the halo mass, $M_\mathrm{vir}$, corresponding to
constant number density, $n(>\!\!M_\mathrm{vir}) = 2 \times
10^{-4}$~Mpc$^{-3}$, at each redshift.  Halos selected at constant
number density grow in mass by roughly a factor of 10, similar to the
growth in stellar mass for galaxies we observe at these redshifts (figure~\ref{fig:zvmass}).
}\label{fig:halomass}
\end{figure}

We studied how well selecting galaxies at constant number density
tracks the descendants and progenitors of galaxies at different
redshifts  by exploring the halo-merger trees and mass functions from
the Millennium simulation \citep[\eg,][]{spri05a}.  This allows us to
study halo growth via mergers and accretion in a physically
motivated simulation that reproduces observational clustering
statistics.  We selected dark-matter haloes with mass,
$M_\mathrm{vir}$, such that $n(>\!\!M_\mathrm{vir},z) = 2 \times
10^{-4}$~Mpc$^{-3}$, for redshifts $3 < z < 7.3$.
Figure~\ref{fig:halomass} shows that the mass corresponding to this
constant number density increases by more than an order of magnitude
over this redshift range (similar to the growth in stellar mass in
galaxies we observe in figure~\ref{fig:zvmass}).    

\begin{figure}
\begin{center}
\includegraphics[width=80mm]{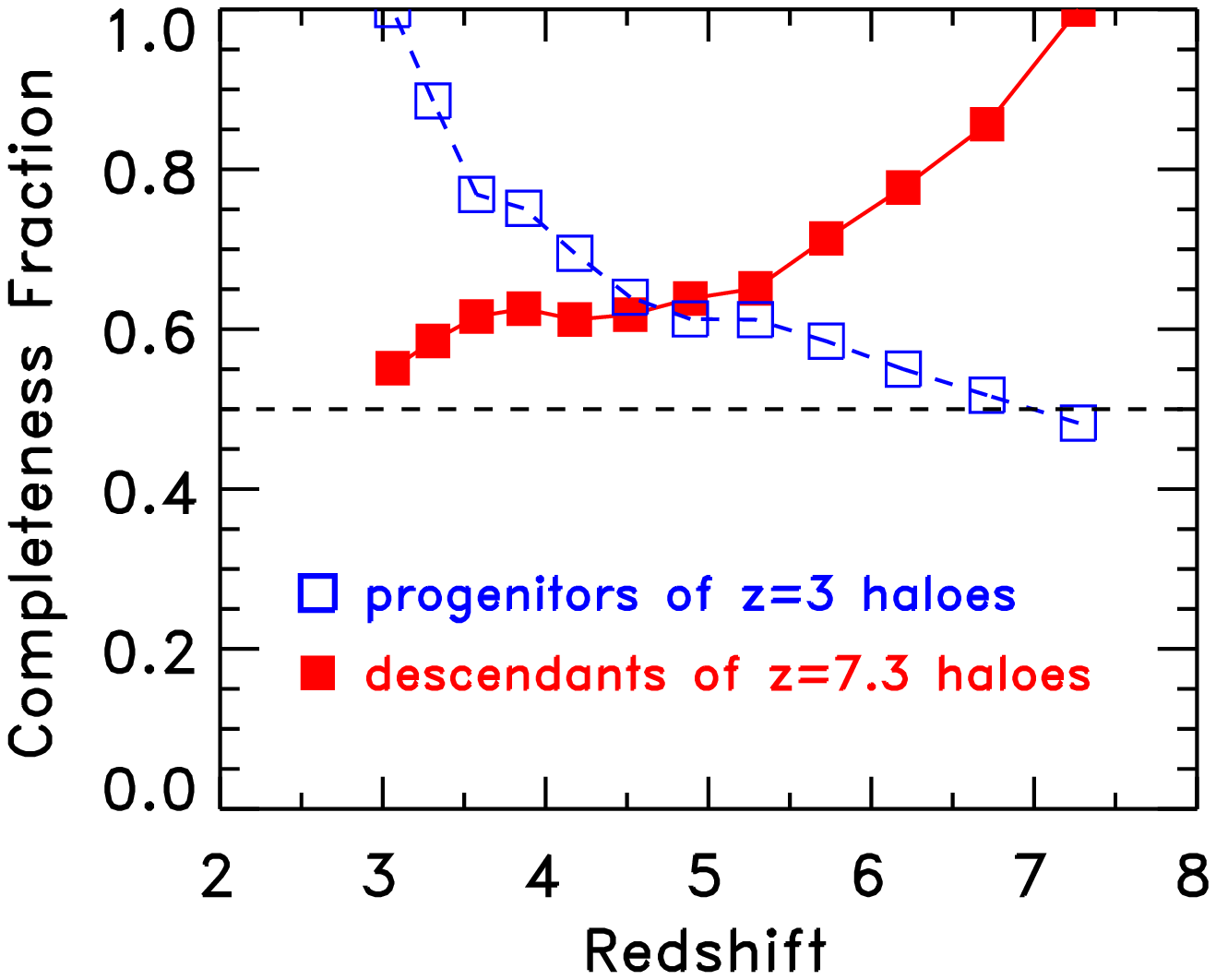}
\includegraphics[width=80mm]{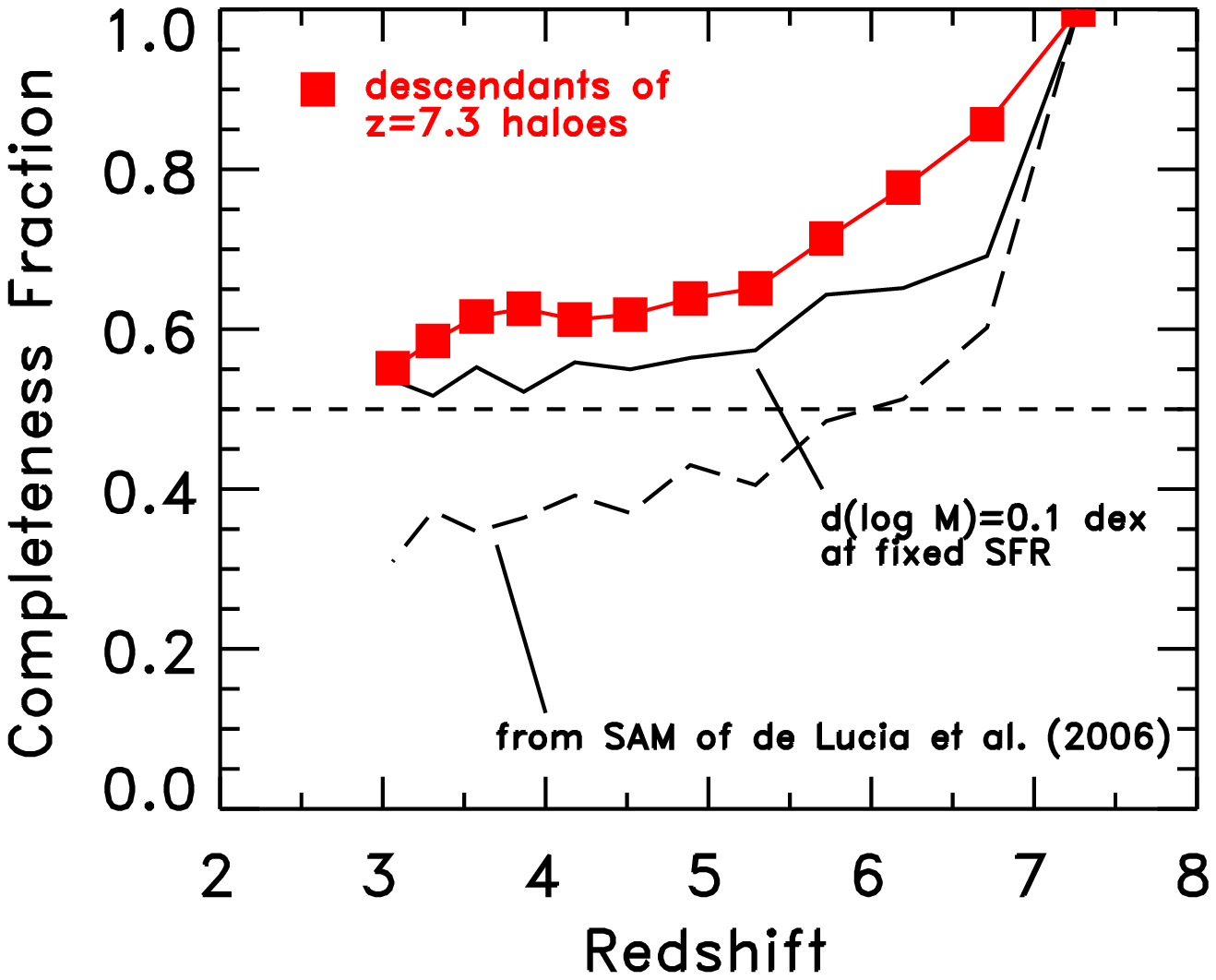}
\end{center}
 \caption{ The top panel shows the completeness of the descendants and
progenitors of dark matter halos at different redshifts.   The filled
squares show the descendant completeness, defined as the fraction of
haloes selected at redshift $z$ with mass such that
$n(>\!\!M_\mathrm{vir}) = 2 \times 10^{-4}$~Mpc$^{-3}$ and a
progenitor at $z=7.3$ selected at this number density.  The
open squares and dashed curve shows the completeness for the
progenitors of haloes selected at constant number density at $z=3$.
The bottom panel shows how the completeness changes for galaxies
selected at fixed SFR with $n(>\!\!\Psi) = 2 \times
10^{-4}$~Mpc$^{-3}$.   The filled squares show the completeness for
the halos as in the top 
panel.  The dashed curve shows the completeness for the SAM of
\citet{delucia06} using the Millennium simulation.  The solid curve
shows the completeness assuming that the scatter in halo mass is
$d(\log M_\mathrm{vir}) = 0.1$~dex at fixed SFR, as predicted by the
models of \citet{dutton10} and \citet{finl10}.
}\label{fig:contcomp}
\end{figure}

%
%
We tracked the descendants of the halos selected in the Millennium
simulation that have halo mass such that $n(>\!\!M_\mathrm{vir},z)$ =
$2\times 10^{-4}$~Mpc$^{-3}$ at $z < 7.3$ and have a progenitor at
$z=7.3$ with mass $M_\mathrm{vir}$ such that it has the same number
density.    The top panel of figure~\ref{fig:contcomp} shows the
``completeness'' fraction, which is defined as the ratio of the number
of the descendants of the galaxies that reside in haloes with mass
that satisfies $n(>\!\!M_\mathrm{vir}) = 2 \times 10^{-4}$~Mpc$^{-3}$
at lower redshift $z$ compared to the number at $z=7.3$.  The
completeness fraction is related to the ``contamination'' fraction,
which is the fraction of those haloes at redshift $z$ with mass such
that $n(>\!\!M_\mathrm{vir}) = 2 \times 10^{-4}$~Mpc$^{-3}$ but
lacking a progenitor with mass such that $n(>\!\!M_\mathrm{vir}) = 2
\times 10^{-4}$~Mpc$^{-3}$ at $z=7.3$.  Because the number density is
constant, the completeness and contamination are related by $(1 -
\mathrm{completeness}) = \mathrm{contamination}$.  

The completeness fraction drops from unity at $z \approx 7.3$ to
approximately 0.6--0.7 for $3 < z < 6$. This implies that selecting
halos at  this number density identifies the majority of the
lower-redshift descendants of objects selected at higher redshift.
The haloes whose descendants are ``missing''  have masses
$<\!\!0.5$~dex below the mass limit for this number density.
Arguably if we allowed an evolving number density criterion, then we
would select more of the halo descendants, increasing the completeness
fraction.  However, this would come at the cost of increasing the
contamination fraction.  Similarly, the haloes that \textit{contaminate} the
sample have masses at $z=7.3$ within $<\!\!0.5$~dex of the mass
limit for the number density selection.  Because these
contaminants have masses close to the selection threshold, we
expect they will have similar evolutionary paths and SFHs as the
galaxies selected at constant number density \citep[see \S~3 and
discussion in][]{finl10}. 

Figure~\ref{fig:contcomp} also shows the completeness of the
\textit{progenitors} of haloes selected at $z=3$ with
$n(>\!\!M_\mathrm{vir}) = 2\times 10^{-4}$~Mpc$^{-3}$.    The
conclusions are similar as those for the descendants of the $z=7.3$
haloes.  The progenitor completeness becomes lower at higher redshifts
compared to the descendant completeness, as the progenitors of $z=3$
haloes become slightly less massive than the required value for the
number density selection.     We conclude that selecting halos at
constant number density traces the majority of the descendants and
progenitors of galaxy halos at $3 < z < 7.3$, and therefore provide
samples to study the direct evolution in these populations.  

 However, thus far we have discussed the completeness in the
descendants and progenitors of only the dark matter haloes.  To make
the connection to observed galaxies, we make the \textit{ansatz} that
there is zero scatter between galaxy mass, galaxy luminosity, and halo
mass, 
\begin{equation}\label{eqn:uv2dm}
n_\mathrm{gal}(>\!\!L_\mathrm{UV}) = n_\mathrm{gal}(>\!\!M^\ast) = n_\mathrm{halo}(>\!\!M_\mathrm{vir}). 
\end{equation} 
The true scatter in equation~\ref{eqn:uv2dm} is very poorly known
theoretically, with values ranging from $d(\log M_\mathrm{vir}) =
0.05-0.1$~dex at fixed SFR \citep[\eg,][]{dutton10,finl10} to 0.3 dex
\citep[\eg,][]{delucia06,bertone07}.    Our assumption of zero scatter
is the simplest assumption possible.  It allows us to study the
evolution of galaxies at constant number density without reliance on
theoretical expectations, and it allows the straightforward comparison
to theoretical predictions for simulated galaxies selected in the same
way.   Moreover, this assumption is justified by observations that
find a tight relationship between UV luminosity and halo mass
\citep[\eg,][]{giav00,adel05b,kslee06,kslee09}.   Even if there is
appreciable scatter, we find evidence that the SFHs of galaxies may
vary only by a scale factor (see \S~3), consistent with expectations
from simulations  \citep[\eg,][]{finl10}.  Therefore, even if our
assumption of zero scatter between UV luminosity and dark-matter halo
mass is simplistic, the derived SFHs still should be reasonably
accurate.

Nevertheless, a non-zero scatter between UV luminosity and dark matter
halo mass would affect the completeness in the descendants and
progenitors of galaxies selected at constant number density at
different redshifts.  We study this effect using several models of galaxy evolution.   
%
%
For the \citet{delucia06} SAM models we select galaxies at $z=7.3$
with SFR such that $n(>\!\!\Psi) = 2\times 10^{-4}$~Mpc$^{-3}$ and
track the descendants within the Millennium merger-tree at lower
redshift directly.   Because these models have the highest scatter at
fixed SFR, $d\log M_\mathrm{vir} = 0.3$~dex, the completeness in the
descendants selected with constant number density at lower redshift
declines.  The bottom panel of figure~\ref{fig:contcomp} shows that
the completeness drops to 40\% in these cases.  However, we consider
this model to be extreme as it has not been shown to reproduce the UV
luminosity functions at high redshift.  In contrast, the
\citet{dutton10} SAMs and \citet{finl10} simulations have lower
scatter, $d(\log M_\mathrm{vir}) = 0.1$~dex, and these models broadly
reproduce trends in the SFR-mass and UV luminosity functions at high
redshift.  For these models we randomly adjust the mass of the dark matter
halos in the Millennium simulation by the scatter  and reevaluate the
completeness.  Figure~\ref{fig:contcomp} shows that for these models
the completeness in the descendants of $z=7.3$ galaxies remains
$>$50\% for all redshifts $z \ge 3$.  In this case selection at
constant number density selects the majority of the descendants as a
function of redshift and has a negligible impact on our conclusions.     

Therefore, even if there is appreciable scatter in the SFR-halo mass
relation, selection at constant number density still tracks the
descendants of the higher-redshift galaxies, albeit with a higher
contamination fraction.    Furthermore, we find that the majority of
the the descendants that are ``missed'' by this selection as well as
those galaxies that ``contaminate'' this selection (those without a
progenitor selected at higher redshift) have SFRs within a factor of 3
compared to the limit for constant number density.     These galaxies
appear to have evolutionary histories similar to the galaxies selected
at constant number density and scatter into and out of the sample at
each redshift. While this reduces the ability of the selection
technique to track the direct progenitors and descendants of the
galaxy population, it should introduce only a weak affect on the
derived SFH.


We therefore conclude that selection at constant number density
provides relatively fair, homogeneous samples with which to study
galaxy descendants and progenitors.  The conclusions here are similar
to those of \citet{vandokkum10} (see discussion at the beginning of
this section).   Comparing galaxy samples at constant number density
seems preferable to study galaxy SFHs, compared to studies that use
luminosity- or mass-selected galaxies.  This is for the reason that
the latter quantities evolve strongly in time.  Samples selected at
constant luminosity or stellar mass will be heterogeneous  at
different redshifts, potentially obfuscating any evolution in these
quantities themselves. 
%

\section{ON THE STELLAR MASSES DERIVED FROM MODELS WITH RISING SFRS}\label{section:appendixB}

One of the main conclusions from this study is that the SFRs of
galaxies at constant number density increase from $z=8$ to
$z=3$ and that this SFH is consistent with the
evolution of stellar mass in these galaxies.   However, their is a
potential inconsistency with this later arguement as galaxy stellar
masses used to make this comparison were derived from models using
either constant or declining SFRs (see \S~\ref{section:masses}).
\citet{mara10} concluded recently that models with rising SFRs
provide better fits to the  measured photometry for $z\sim 2$
star-forming galaxies, and produce stellar masses larger than
constant or decaying $\tau$ models by factors of $\sim 2-3$ \citep[see
also][]{sklee10}.  Therefore, it is prudent to consider how the
stellar masses of galaxies are affected using models with rising SFRs
compared to those for constant or declining SFRs.

We compiled multiwavelength photometry for galaxies in the GOODS-S
field with high fidelity spectroscopic redshifts $3.0 < z < 6.0$
\citep{cris00,stanway04,vanz08,popesso09,bale10}.  The multiwavelength
photometry was originally selected using the ACS \acsz-band, and
includes aperture matched colours in bandpasses from ACS
$\acsb\acsv\acsi\acsz$, ISAAC $JHK$, and IRAC at 3.6, 4.5, 5.8, and
8.0~\micron.  The ISAAC and IRAC photometry are measured using TFIT
\citep{laid07} using the ACS \acsz\ data as a prior for source
positions.  

We derived stellar masses for this subsample of GOODS-S galaxies
using decaying $\tau$ models, $\Psi \sim \exp(-t/\tau)$, following
\citet{papo01}.    We used the 2007 version of the \citet{bruz03}
models with Chabrier IMF and fixed metallicity of 0.2~\zsol.   We
allowed the stellar population age to vary from $10^7 - 2 \times
10^{10}$ yr, with $\tau$ in the range $10^6 - 10^{10}$ and $\tau =
\infinity$ (equivalent to a constant SFR), with dust attenuation
following \citet{calz00} ranging from $A(V) = 0.0 - 2.8$~mag.     We
then derived stellar masses using the same models, but where the SFRs
rise with time following equation~\ref{eqn:sfrvz}, and where we fix
the age ($t$) to be the lookback time from the galaxy redshift $z$ to
the fiducial formation redshift $z_f = 11$.   We allow for the same
range of dust attenuation, and we allow the overall normalization of
the model to vary in order to fit each galaxy because the evidence
suggests that the SFHs of galaxies differ only by a factor that scales
with mass \citep[see \S~3, and][]{finl10}.     We note that this SFH
is the cosmological average of galaxies selected at constant number
density.  In reality individual galaxies will have unique SFHs that
likely depart from this model.  The point here is not to obtain
necessarily the best representative fit to the photometry of each
galaxy, but rather to understand how models with rising SFRs affect
the derived stellar masses.

\begin{figure}
\begin{center}
\includegraphics[width=70mm]{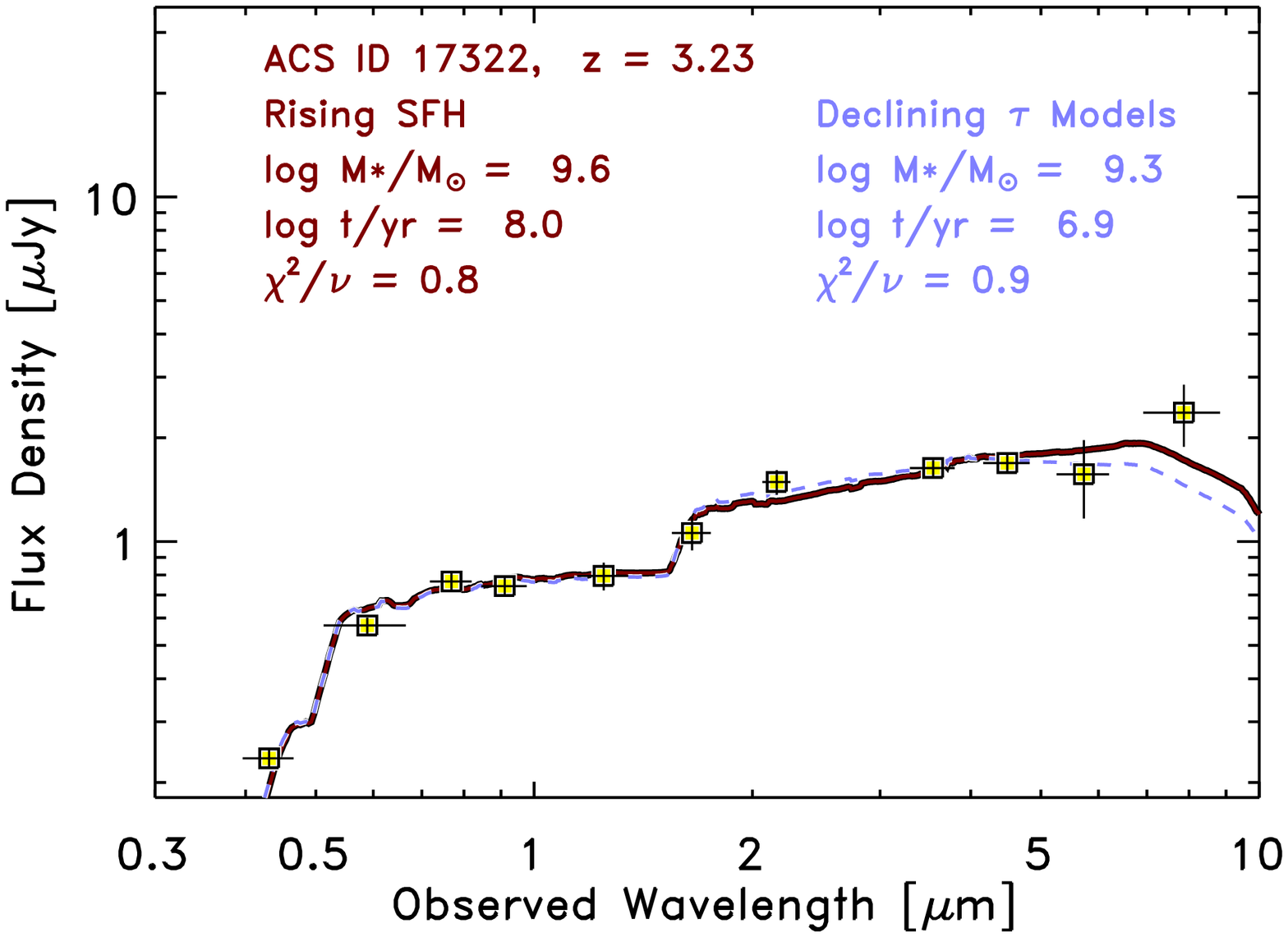}
\includegraphics[width=70mm]{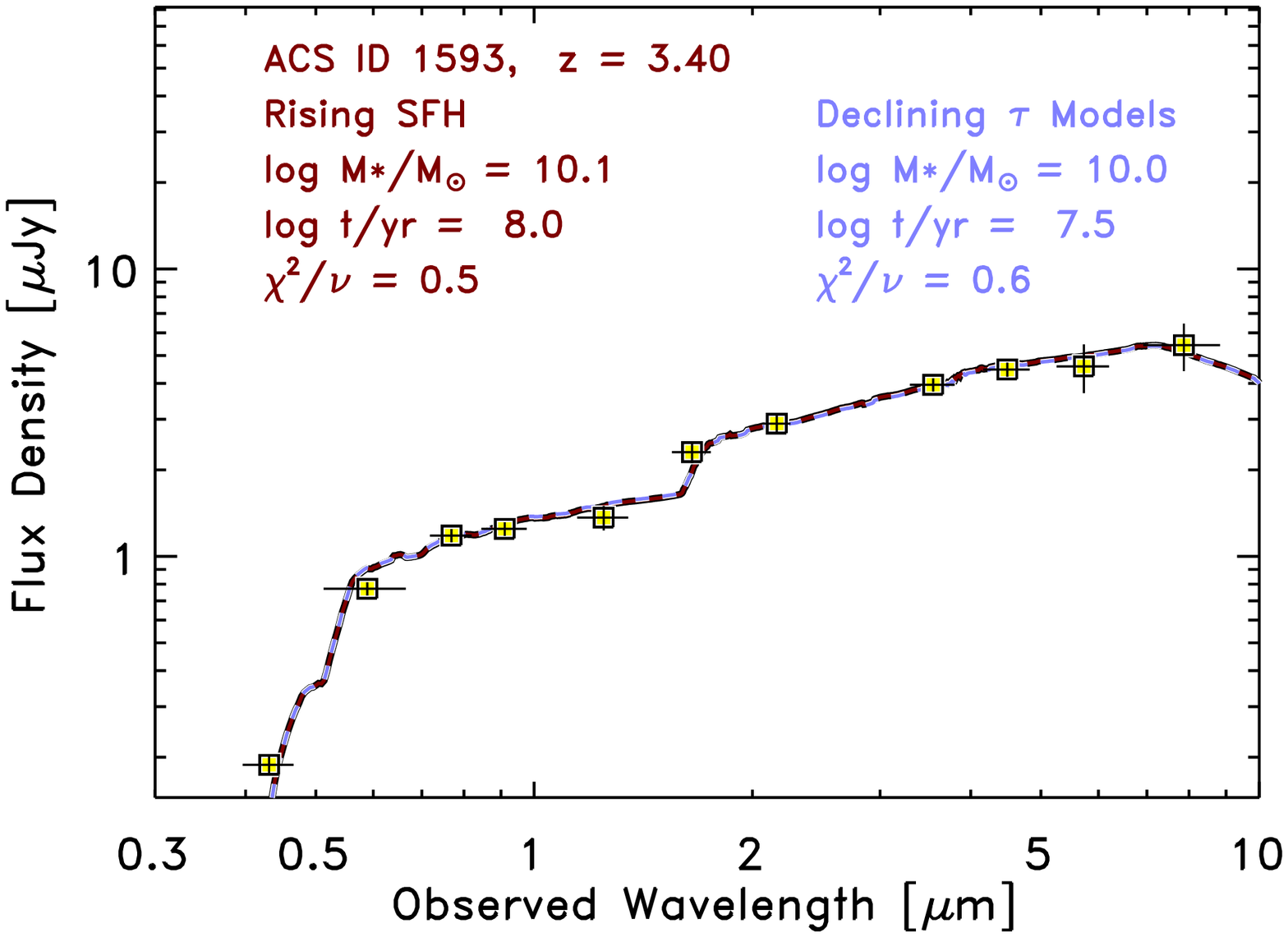}
\includegraphics[width=70mm]{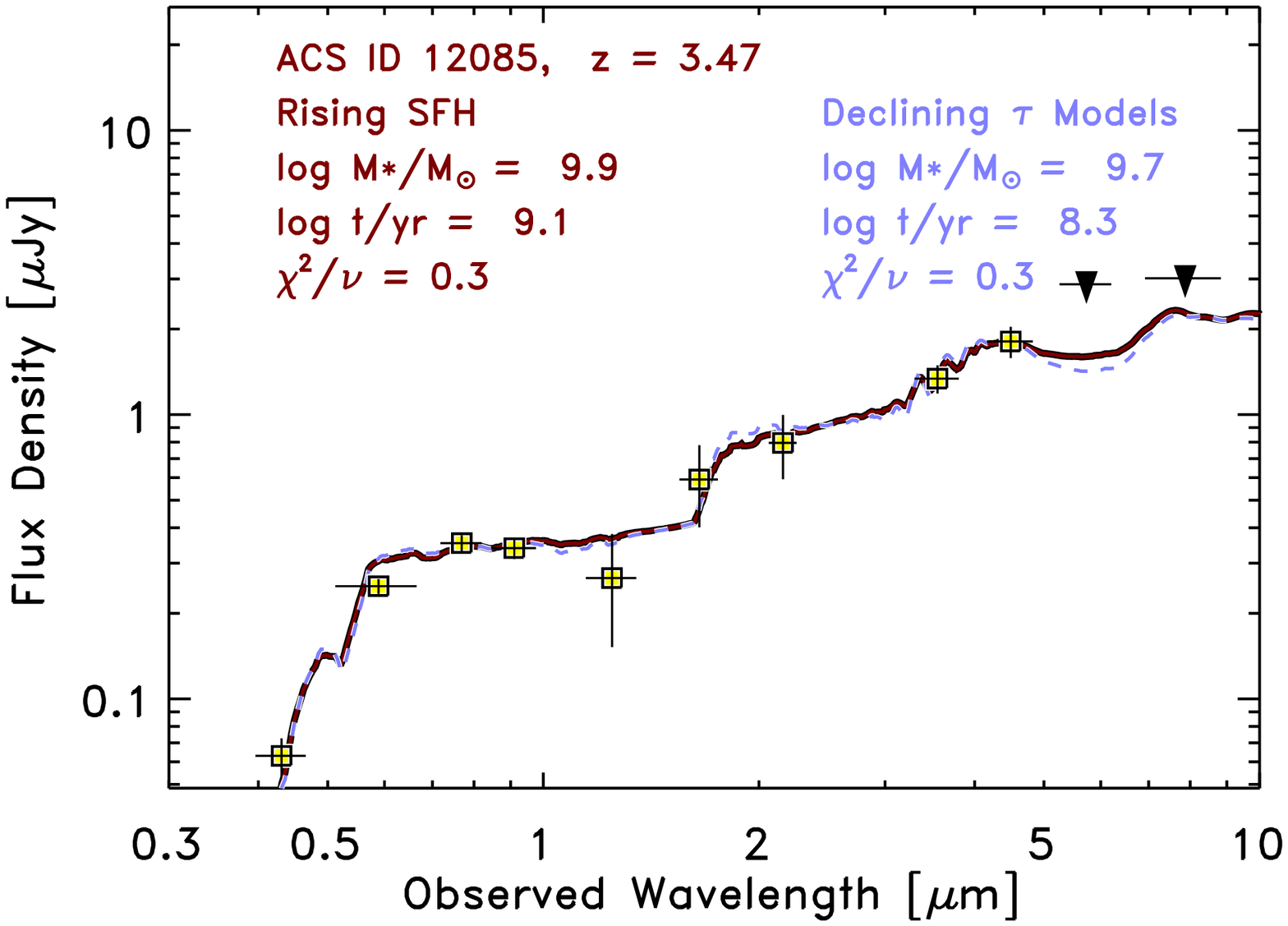}
\end{center}
\caption{Fits to the broad--band photometry for several galaxies from
  the spectroscopic GOODS-S subsample.   Each panel shows the results
  for an individual galaxy.  The ACS, ISAAC, and IRAC photometry are
  indicated by yellow squares.  The curves show the best-fit models
  for models with rising SFHs (brownish-red curves) and declining SFHs
  (light-blue curves) with best-fit parameters inset in each panel.
  The full range of model parameters used are listed in the text.   In
  general, models with rising SFHs yield higher stellar masses and
  older ages, but both models have best-fit reduced $\chi^2$ that are statistically generally indistinguishable. }\label{fig:sedfits}
\end{figure}

Figure~\ref{fig:sedfits} shows several examples of galaxies with
stellar masses that would fall in a sample selected at constant number
density, $n = 2 \times 10^{-4}$~Mpc$^{-3}$.   The curves in each panel
show the rising SFH and declining SFH models with the ``best-fit''
parameter values from the model with the minimum reduced $\chi^2$ value with
parameters listed.    In all cases the stellar population ages and
stellar masses have larger values for models with rising SFHs.   The
minimum reduced $\chi^2$ are close in all cases.  This is a result of the fact
that the SFH is poorly constrained by modeling the broad-band
photometry of galaxies \citep[\eg,][]{papo01}.  However, in general
the models with rising SFHs provide fits that are statistically
indistinguishable  as models with declining models, which is
noteworthy given that here we have tested only a single rising SFH
(with $\Psi(t) \sim t^{1.7}$) compared to the declining models, which
spanned the full range of $e$--folding time-scales (see above).
Finally, we remind the reader that we expect models with rising SFHs
to represent \textit{average} galaxies with ongoing star-formation.
Realistically, galaxies have individual, stochastic SFHs.   In
particular, at lower redshifts, $z \lsim 3$, galaxies will exhibit
``suppressed'' SFRs, and these are consistent with declining SFHs.
The rising SFHs advocated here corresponds to the average evolution of
the star-forming galaxy population at early times ($3 \lsim z \lsim
8$).  

\begin{figure}
\includegraphics[width=84mm]{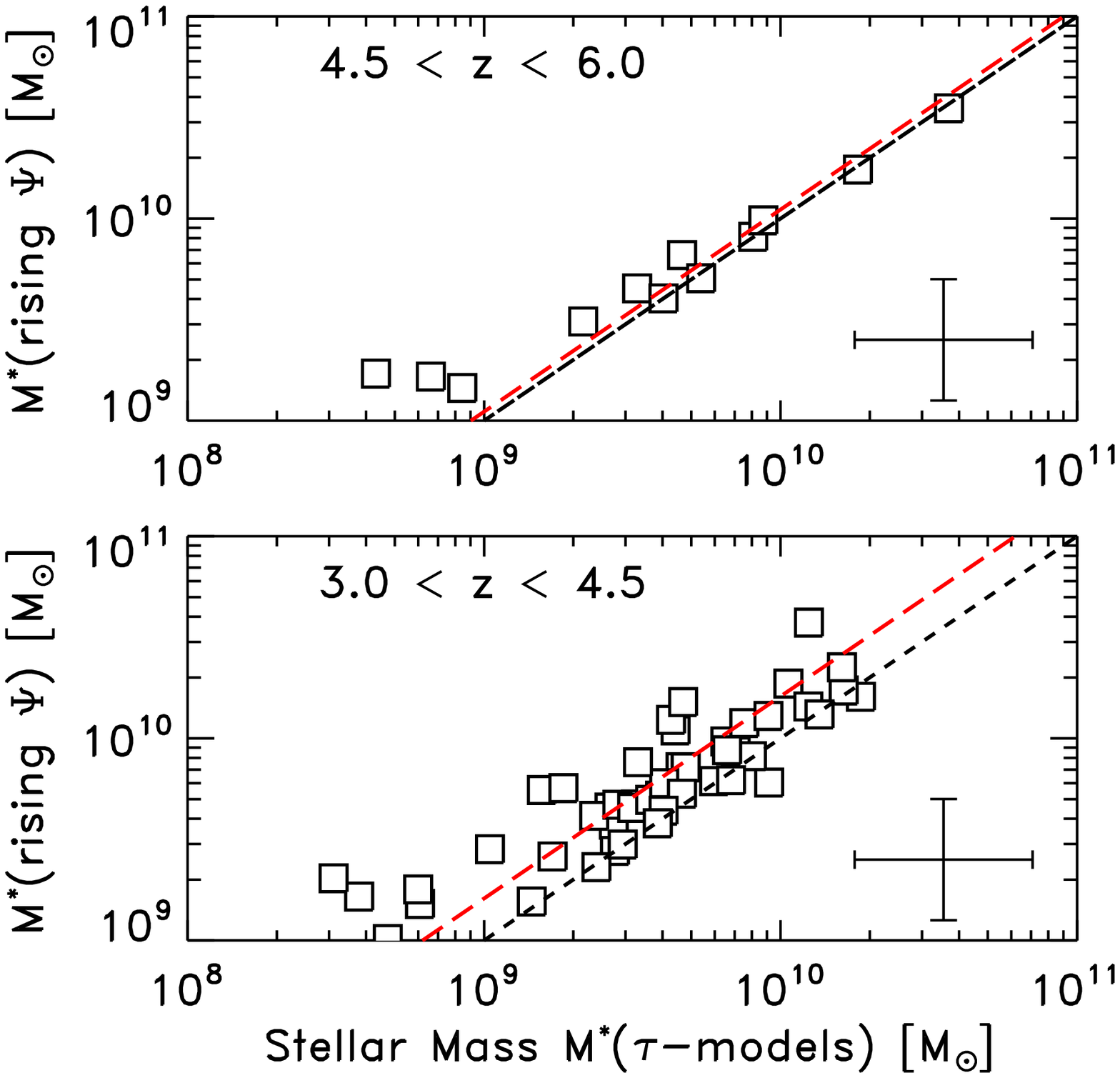}
\caption{Stellar masses derived from decaying $\tau$ models where
$\Psi \sim \exp(-t / \tau)$ versus those derived using the rising SFH
derived here (equation~\ref{eqn:sfrvz}).   The panels show galaxies
with spectroscopic redshifts from GOODS-S in two redshift bins of $3 <
z < 4.5$ and $4.5 < z < 6$. Representative error bars are shown.  The
black, short-dashed lines show the unity relations.  Stellar masses
derived with rising SFHs generally produce higher stellar masses than
decaying $\tau$ models.    The increase in stellar mass generally
increases with time.  The stellar masses derived from models with
rising SFRs for galaxies at $3 < z < 4.5$ are a factor of 1.6 higher
(indicated by the red, long-dashed line, bottom panel), while the
offset is smaller (factor of 1.1, red long-dashed line, top panel)
for galaxies at $4.5 < z < 6.0$. }\label{fig:massmass}
\end{figure}

The results of our fitting analysis show that stellar masses derived
with rising SFRs generally produce higher stellar masses than decaying
$\tau$ models, and that the increase in the stellar masses generally
increases with the passage of time.  Figure~\ref{fig:massmass}
compares the stellar masses derived from the decaying $\tau$ models to
those derived using the rising SFRs, divided into two subsamples of
redshift, $3.0 < z < 4.5$ and $4.5 < z < 6.0$.  The stellar masses
derived from rising SFHs for galaxies at $3 < z < 4.5$ are a factor of
1.6 higher on average, while the offset is smaller (a factor of 1.1)
for galaxies at $4.5 < z < 6.0$.   Therefore, if the model SFH posited
in \S~3 is correct, then the stellar masses of star-forming galaxies
at $z \sim 3-4$ may need to be revised upwards by factor  of order 2,
similar to the conclusions of \citet{mara10} \citep[although see][who
find that rising SFHs overpredict stellar masses of galaxies in
simulations by factors of up to $\sim$2]{sklee10}.  The masses of
galaxies at $z \gsim 4$ are mostly unchanged.  Moreover, this increase
in stellar mass from rising star-formation histories for galaxies at
$z \lsim 4$ is approximately the amount needed to reconcile the offset
between the measured SFRs and stellar masses
(figure~\ref{fig:zvmass}).   If the stellar masses of the UV-selected
galaxies are higher by 0.2-0.3~dex, it would adjust them closer to the
derived SFH.  
%


\bsp

\label{lastpage}

\end{document}